\documentclass[a4paper]{aa}
\usepackage{graphics,rotating}

\newcommand{\apj}[1]{{ApJ }{ #1}}
\newcommand{\apjss}[1]{{ApJS }{ #1}}
\newcommand{\aj}[1]{{AJ }{ #1}}
\newcommand{\aea}[1]{{A\&A }{ #1}}
\newcommand{\aeass}[1]{{A\&AS }{ #1}}
\newcommand{\mnras}[1]{{MNRAS }{ #1}}

\newcommand{\araa}[1]{{ARA\&A }{ #1}}

\newcommand{\kms}{$\rm{km\,s^{-1}}$} 
\newcommand{\mic}{$\mu$m}

\newcommand{\irnulvier}{IRAS\,04296+3429}
\newcommand{\irnulvijf}{IRAS\,05341+0852}
\newcommand{\irnulzeven}{IRAS\,07134+1005}
\newcommand{\ireennegen}{IRAS\,19500$-$1709}
\newcommand{\irtweetwee}{IRAS\,22223+4327}
\newcommand{\irtweedrie}{IRAS\,23304+6147}

\begin{document}
\thesaurus{06(08.01.1; 08.03.1; 08.05.3; 08.16.4)}
\title{A homogeneous study of the s-process in 
the 21 \mic\, carbon-rich post-AGB objects.
\thanks{based on observations collected at the European Southern
 Observatory in Chile (61.E-0426), and at Roque de los Muchachos at La 
Palma Spain}}

\author{Hans Van Winckel\inst{1}\thanks{Postdoctoral fellow of the Fund
for Scientific Research, Flanders} \and Maarten Reyniers\inst{1}\thanks{Scientific 
researcher of the Fund for Scientific Research, Flanders}}

\offprints{Hans Van Winckel, e-mail: Hans.VanWinckel@ster.kuleuven.ac.be.}

\institute{Instituut voor Sterrenkunde, K.U.Leuven, Celestijnenlaan 200B, B-3001 Leuven,
Belgium}

\date{Received 2 July 1999 / Accepted 21 Sept. 1999}

\titlerunning{s-process in post-AGB stars}
\authorrunning{H. Van Winckel \& M. Reyniers}
\maketitle
\begin{abstract}

We present in this paper a homogeneous photospheric abundance study, on the basis of the analysis
of high resolution optical spectra, of six post-AGB objects 
displaying a 21\mic\, circumstellar dust feature in their IR spectrum.
The F-G spectral type of the 21\mic\, stars make that a large range of
elements including a wide variety of s-process elements, can be studied by their atomic lines.
The high C/O-ratios together with the large overabundance of s-process
elements prove that the objects are descendants of field carbon stars.
We discuss in detail the s-process abundance distribution displayed by these
21\mic\, stars and conclude that the 3rd dredge-up efficiency is closely related
to the strength of the integrated neutron irradiation. 
The expected anti-correlation
of the neutron irradiation with metallicity, on the other hand, contains a 
large {\sl intrinsic} scatter. Finally we compare our results with other intrinsic
and extrinsic s-process enriched objects and conclude that the post-AGB stars offer very 
useful complementary data to constrain the evolutionary models of AGB nucleosynthesis 
and dredge-up processes.


\end{abstract}


\section{Introduction}

A sub-class of post-AGB objects display in their IR
spectrum an emission feature at 21\mic, first identified 
on IRAS LRS spectra by Kwok et al. (1989). It was realised 
that the feature only occurs in spectra of C-rich post-AGB 
stars as evidenced by the presence of PAH-emission in the mid-IR spectra
and often strong HCN measurements. The carrier(s) of the 21\mic\,
feature is (are) still unidentified. Its excitation and/or formation is
limited to a short evolutionary phase since the feature
is not observed in IR spectra of AGB stars nor in those of planetary nebulae 
(e.g. Kwok et al. 1999; Volk et al. 1999).
The carbon rich nature of the circumstellar environment of
these objects was corroborated by the detection of 
C$_{2}$, C$_{3}$ and CN bands in the optical
spectra (Hrivnak 1995). Detailed analysis of high-resolution spectra of
these bands indicates that they
are formed in a relatively narrow shell in the circumstellar 
envelope (Bakker et al. 1996; 1997). 

Several studies of the photospheric composition confirmed 
the post-3rd dredge-up character of
individual 21\mic\ stars: they do not only
display a high photospheric C/O ratio but also a
large overabundance of s-process elements (Van Winckel 
1999 and references therein). The colour-temperature of the IR-excess, the 
high luminosity and actual spectral type together with the chemical 
composition of both the photosphere and circumstellar material are all 
observational evidence that the 21\mic\ stars are {\sl post-carbon stars}.

Although in recent years the efforts and results on the theoretical
modelling of the structural and chemical evolutionary AGB models are impressive (see e.g.
Straniero et al. 1995; Frost \& Lattanzio 1996; Herwig et al., 1997; Mowlavi et al. 1998; Mowlavi 1999; 
Langer et al. 1999) the calculations are not
only dependent on fundamental parameters like initial mass and metallicity,
but also on much less constrained or even free parameters 
like mass-loss history and geometry, details of the dredge-up phenomena 
and thermal pulse nucleosynthesis, numerical treatment of convection,
interpulse phase at which the post-AGB phase starts, engulfed proton profile into the intershell,
etc. 

There is now general agreement that the neutrons, needed
for the s-process nucleosynthesis in AGB stars, are mainly produced by the $^{13}$C($\alpha$,n)$^{16}$O
reaction with a possible contribution from the minor activation of the reaction $^{22}$Ne($\alpha$,n)$^{25}$Mg 
(e.g. Smith \& Lambert 1986; Jorissen \& Arnould 1989; Busso et al. 1992; Gallino et al. 1998;
Mowlavi et al. 1998). 
The $^{13}$C available from the CNO hydrogen burning is insufficient to explain observed s-process
abundances (e.g. Mowlavi et al. 1998) so one has to invoke a primary source
of $^{13}$C. This $^{13}$C pocket is naturally build up after
engulfment of free protons into the $^{12}$C-rich intershell. The engulfed proton profile is, however, 
still a free parameter in the predictions. Straniero et al. (1995) proposed that
the s-process synthesis occurs during the intershell in a radiative environment and they computed that
the neutron irradiation varies within the radiative intershell depending on the engulfed proton profile.

The observational data guiding the theoretical chemical 
evolutionary models come from the analysis of intrinsic AGB stars of the M-MS-S-SC-C star sequence, 
(e.g. Lambert et al. 1995 and references therein) which is thought to 
reflect, at least partly, the evolution on the AGB of single stars towards an increasing C/O ratio.
In addition, the enrichment in $^{12}$C seems to correlate well with an increase in s-process
abundances (e.g Smith \& Lambert 1990). The study of the abundance distribution of different 
s-process elements is therefore a very useful tool to characterise the internal nucleosynthesis 
during and in between thermal pulses and to confine models of chemical evolution on the AGB. 
The strong molecular opacity and sometimes unstable photospheres of AGB stars make a quantitative 
analysis, however, difficult and some species even untraceable. 

A second group of objects with excess abundances of s-process elements is the 
Ba-stargroup. In the rest of the paper we use the Ba-stargroup as a generic term for all objects for which 
the excess is acquired by mass-transfer from the companion which is now a white 
dwarf. It includes strong and mild Ba giants and Ba dwarfs, CH giant and subgiants, extrinsic S stars and 
since recently 
also yellow symbiotics (e.g. review by Jorissen 1999). The characteristics of the s-process excesses
in these stars reflect the internal chemical enrichment of the 
companion, and the abundances may have been diluted by mixing of unprocessed material during the evolution
of the gainer. The S-type stars in this group are normally labeled extrinsic S-stars and are
characterised by the lack of Tc in their atmospheres (e.g. Jorissen et al. 1993) or
their lower luminosity than intrinsic AGB S-stars (Van Eck et al. 1998). 

In this paper, we analyse another group of intrinsic s-process enriched objects, namely the 21\mic\, stars.
With their F to G spectral types, a wide variety of chemical species can be 
studied quantitatively by using photospheric atomic lines. This makes them ideal complementary 
sources to constrain the 3rd dredge-up nucleosynthesis and dredge-up models.

The chemical composition of individual 21\mic\, objects
is already reported in the literature (Klochkova 1995;
Za\v{c}s et al. 1995; Van Winckel et al. 1996a; Reddy et al. 1997; 
Decin et al. 1998, Klochkova et al. 1999). Since the different authors
use different line-lists, spectral resolution and
atomic data, it is often difficult to compare the results quantitatively.
We therefore present in this paper a homogeneous chemical 
study of a sample of 21\mic\ objects using the same
spectral resolution, line-lists and atomic data for all 6 programme stars.
We focus specifically on the different s-process elements and
the integrated neutron irradiation. 
In Sect. 2 we give an overview on the data 
obtained and sketch the reduction method, while the details of the chemical analysis are given
in Sect. 3. In Sect. 4 a synopsis of our abundance analysis is presented 
proving that the 21\mic\ stars show clear chemical evidence for their
post-AGB character. Sect. 5 is devoted to the discussion of the results of the 
individual objects and in Sect. 6 we focus on the heavy-element nucleosynthesis and
the determination of the neutron exposure. Our findings on the post-AGB stars are then
compared in Sect. 7 to the s-process abundances displayed in intrinsic 
AGB stars and extrinsic heavy-element enriched objects with a wide range in 
metallicity. We end with formulating the main conclusions in Sect. 8.


\section{Observations and reduction}

In Table~\ref{tab:program} we list the sample of
21\mic\, stars discussed here together with the
spectral domain observed and signal-to-noise obtained.
Most spectra were obtained using
the Utrecht Echelle Spectrograph (UES) mounted on the
4.2m William Hershel Telescope (WHT) on La Palma, Spain. 
We used the echelle with 31.6 lines/mm and the projected slit width 
was 1.1'' on the sky yielding a resolution of around 
R = $\lambda$/$\delta$$\lambda$ $\sim$ 50 000) depending on the wavelength. 

\begin{table*}
\caption{The programme stars: magnitude, coordinates, log of the observations and radial velocities.\label{tab:program}}
\begin{tabular}{|l|l|r|rr|rr|}
\hline
IRAS       & Other           &\multicolumn{1}{|c|}{Visual}&\multicolumn{2}{|c|}{Equatorial}&\multicolumn{2}{|c|}{Galactic} \\
           & name            &\multicolumn{1}{|c|}{magnitude}&\multicolumn{2}{|c|}{coordinates}&\multicolumn{2}{|c|}{coordinates} \\
           &                 &m(v) &$\alpha_{2000}$&$\delta_{2000}$ & l & b\\
\hline
04296+3429  &            & 14.2& 04 32 56.6  & $+$34 36 11    & 166.24 &   $-$9.05  \\
\hline
05341+0852  &            & 12.8& 05 36 54.2  & $+$08 54 10    & 196.19 &  $-$12.14  \\
\hline
07134+1005 & HD\,56126  & 8.3 & 07 16 10.2 & $+$09 59 47  & 206.75 &    +9.99  \\
\hline
19500$-$1709 & HD\,187885 & 9.2 & 19 52 52.7 & $-$17 01 50  &  23.98 &  $-$21.04  \\
\hline
22223+4327 &            & 9.7 & 22 24 30.7 & $+$43 43 03  &  96.75 &  $-$11.56  \\
\hline
23304+6147 &            & 13.1& 23 32 45.0  & $+$62 03 49    & 113.86 &    +0.59  \\
\hline\hline
            & \multicolumn{1}{|c|}{Date and UT} &Telescope+   & Sp. Range. & S/N      & $v_{\rm helio}$ & $v_{\rm LSR}$ \\
            &             &Spectrograph & (nm)       &          &  (\kms)           &  (\kms)        \\
\hline
04296+3429   &  22/2/1994 21:20& WHT+UES  & 556.5-1022   & 80          &   $-$53&   $-$60 \\
\hline
05341+0852   &  08/8/1995 05:15& WHT+UES  & 556.5-1004   & 50       &    24      &     8 \\
\hline
            &                 &          &              & blue : 160 &             &       \\
\raisebox{1.5ex}[0pt]{07134+1005}  & \raisebox{1.5ex}[0pt]{29/9/1998 09:01}&\raisebox{1.5ex}[0pt]{NTT+EMMI} &\raisebox{1.5ex}[0pt]{398-662} &
green : 210  &\raisebox{1.5ex}[0pt]{86} & \raisebox{1.5ex}[0pt]{71} \\
            &  24/2/1992 21:57& WHT+UES  & 535-1040     & 100      &    84       &    70 \\
\hline
19500$-$1709  &  06/8/1995 23:55& WHT+UES  & 364-458      & 140      &    12     &    24 \\
            &  08/8/1995 00:17& WHT+UES  & 453-680      & 240      &    13       &    25 \\
            &  07/8/1995 01:34& WHT+UES  & 551-1004     & 320      &    12       &    24 \\
\hline
22223+4327  &  23/8/1994 01:56& WHT+UES  & 443.5-650.5  & 180      &   $-$42&   $-$30 \\
            &  21/8/1994 22:45& WHT+UES  & 556.5-1004   & 190      &   $-$42&   $-$30 \\
\hline
23304+6147  &  23/8/1994 03:48& WHT+UES  & 443.5-650.5  &  50      &   $-$26&   $-$16 \\
            &  22/8/1994 01:09& WHT+UES  & 556.5-1004   & 110      &   $-$26&   $-$16 \\
\hline
\end{tabular}
\end{table*}

\renewcommand{\irnulvier}{IRAS04296}
\renewcommand{\irnulvijf}{IRAS05341}
\renewcommand{\irnulzeven}{IRAS07134}
\renewcommand{\ireennegen}{IRAS19500}
\renewcommand{\irtweetwee}{IRAS22223}
\renewcommand{\irtweedrie}{IRAS23304}

Some of the UES spectra were kindly provided by Dr. Eric Bakker 
and used by him in his
analysis of the optical circumstellar
molecular absorption bands (Bakker et al. 1996; 1997).
For \irnulzeven\ the 3.5m New Technology Telescope
(NTT) of the European Southern Observatory (ESO) was used in combination 
with the high-resolution echelle mode of EMMI 
using grating number 14. The projected slit width
was 1'' on the sky, yielding a resolution of R $\sim$ 60 000.
Since most objects are heavily reddened, the S/N 
achieved depends strongly on wavelength and of course on 
the specific characteristics of the cross-dispersed spectrographs. The standard 
reduction included bias correction, flat-fielding, cosmic hits identification
and cleaning and background subtraction. 
The extraction was done by a simple mean
over the spatial profile and the wavelength calibration was performed by using
Thorium-Argon lamp measurements. The different orders
do not overlap from  $\sim$ 580\,nm red-wards for the UES
in combination with the 1024$^{2}$ pixel TEK CCD and 
for 742\,nm red-wards for EMMI with the 2048$^{2}$ TEK CCD 
(ESO number 36). We normalised the
spectra by dividing the individual orders 
by a smoothed spline function defined through interactively identified
continuum points. We used the specific echelle
context of MIDAS for the reduction.


\section{Analysis}

\subsection{Atomic data}

Mainly the solar line list of Th\'evenin (1989,1990) was used for the first line 
identification. For specific elements, the line lists of the Vienna Atomic Line
Database (VALD2, http://www.astro.univie.ac.at/~vald/, Kupka et al. 1999) were used, and 
estimated line strengths were computed. Next, the lines stronger than our detection limit 
were identified in the measured spectra.

The oscillator strengths ($\log(gf)$-values) were taken from several sources:
for Fe, we used the critical compilation of Lambert et al. (1996), completed with values
of Blackwell et al. (1980) for Fe\,II. Values for C, N and O were mainly
taken from the Opacity Project (Hibbert et al. 1991, 1993; Bi\'emont et al.
1991);
values for the s-process elements from the Vienna Atomic Line Database (VALD2);
values for the other elements
mainly from Th\'e\-ve\-nin (1989, 1990), but also from 
Wiese et al. (1966),
Reader et al. (1980),
Fuhr et al. (1988),
Martin et al. (1988),
Venn (1995) and
Gonzalez et al. (1997).
The line list can be obtained from the authors upon request.
 
\subsection{Radial velocities}

\begin{table}
\caption{Heliocentric radial velocities of the programme stars, 
chronologically tabulated. The data used for the determination of the 
velocities include CO, and CS rotational line emission 
and optical spectra (opt.). 
The velocities we found are in bold. For these velocities, a detailed time
indication can be found in Table~\ref{tab:program}, together with their
value with respect to the Local Standard of Rest ($v_{\rm LSR}$).\label{tab:vrad}}
 
\begin{tabular}{|l|rrr|r|}
\hline
IRAS       & $v_{\rm helio}$ & method & date         & ref.   \\ 
           &  (\kms)         &        & (mm/yy)      &        \\
\hline
04296+3429 & $-$55 & CO    & 12/88    & 11 \\ 
           & $-$59 & CO    & 01/89    & 7 \\
           &{\bf $-$53} & opt. & 02/94    &     \\
           & $-$56 & opt. & 02/97 & 3 \\
\hline
 
05341+0852  & {\bf 24} & opt.   & 08/95    &      \\
            &  25 & opt.   & 12/96    & 10 \\
            & 26  & CO     & 04/98     & 12  \\
\hline
07134+1005 &  87 & CO      & 01/89    & 7 \\
           & 105 & opt.   & (?)              & 9 \\
           &  82--92 & opt.   & 01/91-04/92 & 5 \\
           &  86 & CO      & 04/90-01/91 & 1 \\
           & {\bf 84$^a$} & opt.   & 02/92       &     \\
           &  91 & opt.   & 01/93 & 2 \\
           &  86 & opt.   & 12/93\&02/94 & 8 \\
           & {\bf 86$^b$} & opt.   & 09/98       &     \\
\hline
19500$-$1709 &  13 & CO      & 03/86  &  6 \\
             &  13 & CO      & 07/87  &  4 \\
             &  13 & CO      & 07/90  &  7 \\
             &  12 & CO    & 04/90--01/91 & 1 \\
             & {\bf 12} & opt. & 08/95    & \\
 
\hline
22223+4327  &  $-$37& CO    & 01/88  & 6 \\
            &  $-$42& CO    & 01/89  & 7 \\
            & {\bf $-$42}& opt. & 08/94  & \\
\hline
23304+6147  &  $-$26& CO      & 12/88  & 11 \\
            &  $-$26& CO      & 01/88  & 6 \\
            &  $-$26& CS      & 01/89  & 7 \\
            & {\bf $-$26}& opt.    & 08/94  & \\
\hline
\end{tabular}
\\ (a) UES spectra (1992), (b) EMMI spectra (1998)
\\
ref.
(1) Bujarrabal et al. 1992
(2) Klochkova 1995
(3) Klochkova et al. 1999
(4) Knapp et al. 1989
(5) L\`ebre et al. 1996
(6) Likkel et al. 1991
(7) Omont et al. 1993
(8) Oudmaijer \& Bakker 1994
(9) Parthasarathy et al. 1992
(10) Reddy et al. 1997
(11) Woodsworth et al. 1990
(12) Hrivnak \& Kwok 1999
\end{table}

The heliocentric radial velocities of the objects are given in Table~\ref{tab:program} with
an internal accuracy of $\sim$1 km s$^{-1}$.
Comparison with published velocities reveals no significant variability in most objects (see Table~\ref{tab:vrad}). 
For \irnulvier, an accurate velocity is hampered by our lower signal-to-noise. 
\ireennegen\ and \irtweedrie\ are clearly not variable.
For \irnulvijf\ and \irtweetwee\ we have not enough velocities to compare
with, but non-variability is most likely.
The results of \irnulzeven\ are more difficult to interpret since the star
is a known photometric variable.
L\`ebre et al. (1996) 
report on the results of their spectroscopic monitoring campaign,
which spanned 440 days. They deduced radial velocity variations between
81.73 and 91.79 km s$^{-1}$ for this pulsating star 
but did not find evidence for binary motion. The measurement of Parthasarathy et al. (1992) 
deviates significantly from this range but it is difficult to evaluate since they do not give the 
date of observation. From the reception date
of the article itself, it could have been very well during the monitoring
campaign of L\`ebre et al. 

We can conclude that for none of the stars there is evidence for binarity
from the radial velocities obtained so far.

\subsection{Determination of atmospheric parameters and abundances.\label{subs:deter}}

\begin{table}
\caption{Model parameters of the six programme stars.\label{tab:atmosfeerparam}}
\begin{center}
\begin{tabular}{|l|cccc|}
\hline
       & Z      & T$_\mathrm{eff}$ & $\log g$ & $\xi_t$    \\
       &        &     (K)   &          &    (\kms)  \\
\hline
\irnulvier  & $-$0.5  & 7000          & 1.0      & 4.0    \\
\irnulvijf  & $-$1.0  & 6500          & 1.0      & 3.5    \\
\irnulzeven & $-$1.0  & 7250          & 0.5      & 5.0    \\
\ireennegen & $-$0.5  & 8000          & 1.0      & 6.0    \\
\irtweetwee & $-$0.5  & 6500          & 1.0      & 5.5    \\
\irtweedrie & $-$1.0  & 6750          & 0.5      & 3.0    \\
\hline
\end{tabular}
\end{center}
\end{table}

In order to calculate abundances, we used the model atmospheres of Kurucz (1993)
on his CDROM nr. 13. These LTE atmospheres are uniquely determined by the effective
temperature (T$_\mathrm{eff}$), gravity ($\log g$) and overall metallicity (Z). The models
were computed using opacity distribution functions assuming a constant microturbulent
velocity ($\xi_\mathrm{t}$) of 2 \kms.

The model parameters were based solely on the basis of the spectra, more specifically
on the analysis of a set of well measured Fe\,I and Fe\,II lines with accurate atomic data. 
We used Fe for this purpose, not only because of the large number of lines, but also because of the accurate
oscillator strengths. We determined, after several iterations, T$_\mathrm{eff}$ by
demanding the abundances of a Fe-ion to be independent of the lower
excitation potentials; the surface gravity $\log g$ by demanding ionization
equilibrium for Fe\,I and Fe\,II; the microturbulent velocity $\xi_t$,
by forcing the
abundances to be independent of the reduced equivalent 
widths $ \log(W_{\lambda} / \lambda)$. Results of other authors or photometric data were used as a first
guess. The stars in the sample are $\sim$1000\,K too cool to show He\,I-lines, so
we could not obtain He-abundances, except for the hottest star (\ireennegen,
T$_\mathrm{eff}$=8000K). We already reported the detection of a He\,I-line
($\lambda$587.565\,nm) for this star in Van Winckel et al. (1996a).
The obtained parameters are listed in Table~\ref{tab:atmosfeerparam}.

The abundance determinations were made using the Kurucz program WIDTH9. We
refer to Decin et al. (1998) and Van Winckel (1997) for an extensive
error analysis and remind that the uncertainties, assuming LTE, are
mainly induced by uncertain $\log(gf)$ values and an uncertain temperature
determination. Since the model atmospheres of Kurucz are listed with steps of $\Delta T_\mathrm{eff}$ = 250 K
and $\Delta$$log(g)$ = 0.5, that correspond to the uncertainties of the spectroscopic parameter determinations,
we did not interpolate between models.
The possibility to infer an abundance of an ion critically depends on
the presence of useful lines of that ion in the studied spectral range. We only used in this LTE analysis
lines which appear unblended in our spectra and have an equivalent width smaller than 150 m\AA.

\begin{figure}
\resizebox{\hsize}{8.5cm}{\includegraphics{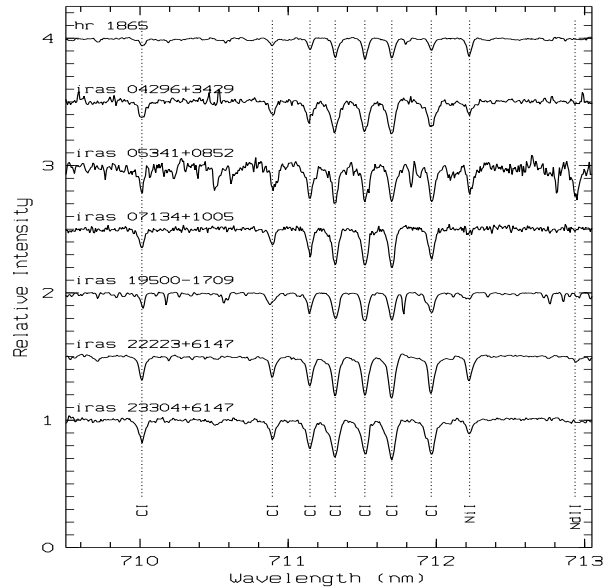}}
\caption{\label{fig:carb}
Sample spectra of the programme stars around the red carbon multiplet. The stars are velocity corrected.
On top, the spectrum of the reference star HR\,1865 is plotted. This massive supergiant (F0Ib) has
similar atmospheric parameters as the programme stars (T$_\mathrm{eff}$=7500\,K,
$\log g$=2.0 and $\xi_t$=3.0 \kms, see Decin et al. (1998)) but obviously no enrichment
of helium burning products. }
\end{figure}

\begin{figure}
\resizebox{\hsize}{8.5cm}{\includegraphics{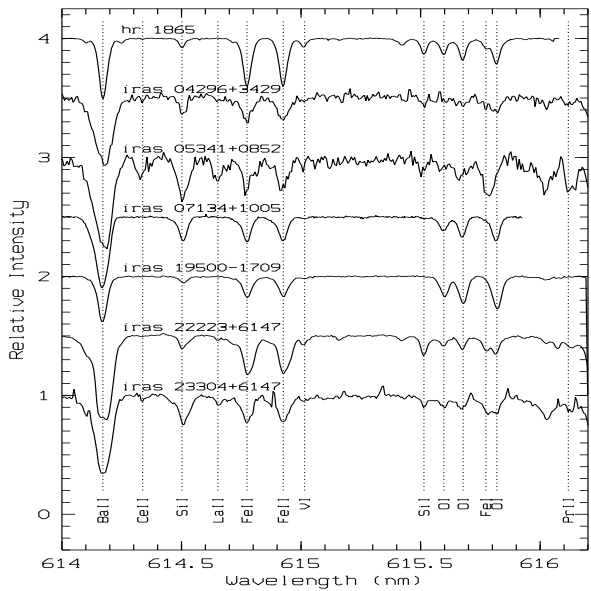}}
\caption{\label{fig:oxy}
Sample spectra of the programme stars and HR\,1865 around the oxygen
triplet at $\lambda$615.6\,nm.
For the cooler stars, a reliable oxygen abundance is rather problematic
to derive, because of the small number of lines: only this multiplet
is available for these stars. Hotter stars (\irnulzeven\ and \ireennegen)
show more O lines. The equivalent width of the BaII-line exceeds our
upper limit of 150\,m\AA\ and can therefore not be used in a barium abundance
calculation. A barium abundance is always difficult to determine because
this element shows only a few very strong saturated lines and weak optical
lines with $W_{\lambda}$$<$150{\rm\,m\AA} are very rare.}

\end{figure}

Lines of neutral carbon (C\,I) are numerous in the spectrum of each
star (see Fig.~\ref{fig:carb}). Combined with accurate $\log(gf)$-values, we
were able to determine intrinsically very consistent C-abundances ($\sigma({\rm C\,I})\le0.14$,
except for \linebreak \irnulvier: $\sigma({\rm C\,I})=0.20$). Useful
nitrogen lines (N\,I) are more difficult to obtain, but there are some weak
lines in the red part of the spectrum; only for \irnulzeven\ we measured
more than 5 weak lines.
The oxygen abundances of the 21\mic\, stars are in most cases more difficult
to obtain. Since the O triplet at $\lambda$777.4 nm is strongly non-LTE
sensitive, weaker lines should be used. For the cooler stars in the sample,
only the high excitation multiplet at 615.6\,nm can be used (Fig.~\ref{fig:oxy}), but since
these lines are heavily blended, we used multiple Gaussian fitting
to estimate their equivalent widths. For the two hotter stars
\irnulzeven\ and \ireennegen, there are more lines in the spectrum, so the
O-abundance is better established.

From sodium (Na, Z=11) to sulphur (S, Z=16), abundances are hard to
obtain due to the lack of lines. Only two different sodium lines were
found in the spectra \linebreak ($\lambda568.265$\,nm and $\lambda568.822$\,nm),
both of them having small equivalent widths ($W_{\lambda}\le26$\,m\AA).
\ireennegen\ and \irtweetwee\ are the only stars that
show aluminium lines which we considered to be good enough to use in
an abundance calculation. Silicon (Si) has more
lines, but it shows for some stars a rather large line-to-line scatter.

For the heavier metals (Ca, Sc, Ti, Cr, Ni), more lines were found.
Neutral calcium (Ca) is found in every star. Also scandium (Sc) and
titanium (Ti) abundances were derived for almost every star, and
this from lines of singly ionised atoms. Chromium (Cr) displays
lines from neutral as well as singly ionised atoms, allowing an extra
check for the ionization equilibrium (see Sect.~\ref{subs:deter}).
Nickel (Ni) is also found in every star, showing neutral lines in the
cooler objects, ionised lines in the hotter ones. Other iron-peak elements found
are vanadium (V), manganese (Mn) and zinc (Zn).

\begin{figure}
\resizebox{8cm}{6cm}{\includegraphics{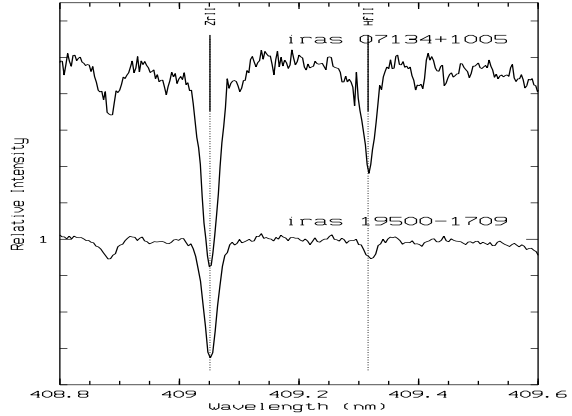}}
\caption{\label{fig:hafnium}
The HfII-line ($\lambda$409.316\,nm) in the blue spectrum of \irnulzeven\ 
and \ireennegen. The derived hafnium abundance is quite high, but blends
or a wrong identification is almost certainly excluded. }
\end{figure}

For the 21\mic\, stars which have F spectral types, the s-process elements
are determined by lines from singly ionised atoms. The strontium (Sr) abundance
is very difficult to determine accurately since the resonance lines at $\lambda$421.5\,nm and
$\lambda$407.7\,nm are heavily saturated and weaker optical lines are not present.
Yttrium (Y) and zirconium (Zr) abundances are more easily obtained. Moreover,
these elements show a large number of lines in the hotter stars of our sample.
An accurate barium (Ba) abundance is, like for Sr, difficult to determine,
since weak optical
lines are absent and only a few very strong saturated lines are present
(for example, see Fig.~\ref{fig:oxy}). Three BaII-lines were found
with $W_{\lambda}<150{\rm\,m\AA}$:
$\lambda$416.600\,nm, $\lambda$585.367\,nm and $\lambda$873.776\,nm.
Lanthanum (La), cerium (Ce) and neodymium (Nd) are easier to study and
we obtained abundances for every star. Praseodymium (Pr) and samarium (Sm) have
less lines and we were not able to determine abundances for these elements for every
star. Also europium (Eu) lines are quite rare;
we detected only three different EuII-lines in our spectra. Finally,
we also detected a Hafnium (Hf, Z=72) line ($\lambda$409.316\,nm)
in the spectrum of \irnulzeven\ and \ireennegen\ (Fig.~\ref{fig:hafnium}).
This line will be discussed in more detail in Sect.~\ref{sec:individobj}.

\begin{figure}
\resizebox{\hsize}{8cm}{\includegraphics{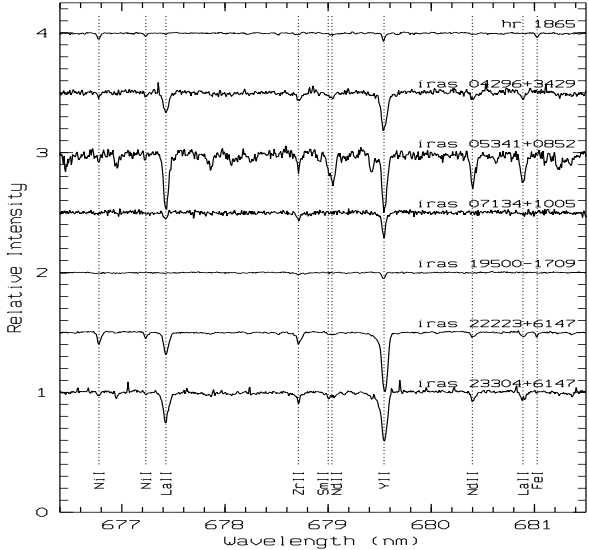}}
\caption{\label{fig:spro}
The region around $\lambda$679\,nm for the six programme stars and HR\,1865.
In this region, both lines of light s-process elements (ls) and
of heavy s-process elements (hs) are present. Comparison with
the normal supergiant HR\,1865 proves that our stars are clearly
s-process enhanced. The ratio of the line strength of the
LaII-line at $\lambda$677.427\,nm and the YII-line at $\lambda$679.541\,nm for each star
gives a first very qualitative idea of the neutron exposure (which is 
discussed in Sect.~\ref{section:neutronexp}). In particular,
from this comparison we can expect a high neutron exposure for \irnulvijf.
}
\end{figure}

\subsection{Hyperfine splitting}

We investigated the influence of hyperfine splitting (hfs) on
some observed lines which are considered to be sensitive for
this effect. Hfs causes extra desaturation but is only important for 
lines of elements with an odd atomic number and strong nucleon-electron interaction.

Hfs will affect weak lines in their profile, but not in their equivalent
width, leaving the derived abundance unchanged. For stronger lines on the
flat part of the curve of growth, however, also the equivalent width will be affected. 
The hfs-influence critically depends on the temperature, since at higher
temperatures the Doppler widths will wash out the effect of hfs on the line-profile.

The Kurucz spectrum synthesis program "Synthe" (Kurucz 1993) was used to
study the influence of hfs. We calculated the difference in equivalent
width of a synthetic line treated with and without hfs-decomposition for the relevant model
atmospheres.
We investigated the following lines which are considered to be among the
most influenced by hfs-splitting: one aluminium ($\lambda$ 394.401\,nm), one scandium
($\lambda$ 660.460\,nm) and two europium lines ($\lambda$ 412.973\,nm and
$\lambda$ 664.506). The hsf-decompositions were obtained from McWilliam et al. (1995) 
and Biehl (1976). 

The influence of hfs on W$_{\lambda}$ of the 
synthetic lines was found to be noticeable, but spectrum synthesis made clear that the
other sources of uncertainty (like continuum placement, errors on the
atmospheric parameters etc.) are much more important in our programme stars. 
The small influence, even on the normally strongly affected Eu 412.973\,nm line may
surprise, but this line is only observed in the hotter stars where it has an
equivalent width of only 32 m\AA\ at most.  For the cooler 21\mic\, objects this line was
not observed but our calculations indicate that this strong line would be heavily affected indeed.

Since line-profile fitting induces other uncertainties and since
the effect on the weak lines used in our analysis is very small, we
did not include hfs profile fitting results in our final table.


\section{Synopsis of the abundance results}

\begin{figure*}
\resizebox{17cm}{10cm}{\includegraphics{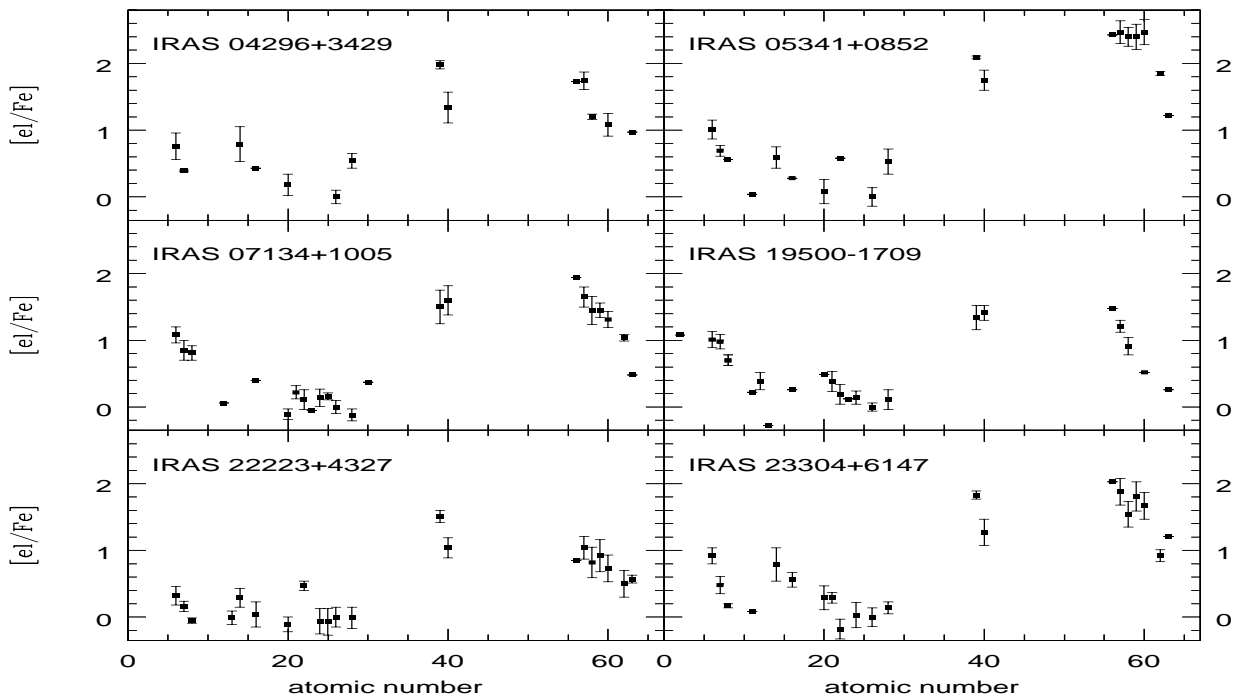}}
\caption{\label{fig:zesgraf}
The abundances of our six programme stars relative to iron [el/Fe].
 }
\end{figure*}

In this section we discuss Table~\ref{tab:synopsis} and Fig.~\ref{fig:zesgraf}, 
which give an overview of the results of our abundance analysis. 
In the last column of Table~\ref{tab:synopsis}, we list
a typical abundance pattern for an unevolved metal deficient star
(e.g. Lambert 1989; Wheeler et al. 1989; Edvardsson et al. 1993; McWilliam 1997) reflecting the
chemical composition of our programme stars at the time of their formation.
Comparing these unprocessed abundances with the observed abundances 
permits us to detect possible
changes in the chemical composition during the stars evolution. Applied
to low-mass post-AGB stars, we expect the chemical signature of the 3rd dredge-up
without activation of hot-bottom burning 
i.e. the enrichment of the stellar photosphere with material exposed to
He-burning during the TP-AGB phase. This signature includes:
carbon and (to a lesser extent) oxygen enrichment, a C/O number ratio higher
than solar (C$_{\odot}$/O$_{\odot}$=0.51) and, above all, s-process enrichment.
All 21\mic\ sources studied here turn out to show almost all 
these enrichments, which make them a group of definite post 3rd dredge-up post-AGB stars.

\begin{table*}
\caption{Synopsis of the abundance results. 
The [$\alpha$/Fe] value is the mean of the [el/Fe] values of the observed
$\alpha$-elements Mg, Si, S, Ca and Ti; [ls/Fe] the mean of
the [el/Fe] values of Sr, Y and Zr; [hs/Fe] the mean of [el/Fe] of
Ba, La, Nd and Sm; [s/Fe] the mean of the former 7 s-process elements.\label{tab:synopsis}}
\begin{center}
\begin{tabular}{|l|l|rrrrrr|r|}
\hline

 &IRAS      & 04296  & 05341  & 07134  & 19500  & 22223  & 23304  & subdwarf \\

\hline
metallicity
 &[Fe/H]    & $-$0.6 & $-$0.8 & $-$1.0 & $-$0.6 & $-$0.3 & $-$0.8 & $-$0.7   \\

\hline
CNO
 &[C/Fe]    & 0.8    & 1.0    & 1.1    & 1.0    & 0.3    & 0.9    & 0.0 \\
abundance
 &[N/Fe]    & 0.4    & 0.7    & 0.9    & 1.0    & 0.2    & 0.5    & 0.0 \\

 &[O/Fe]    &  /     & 0.6    & 0.8    & 0.7    & $-$0.1  & 0.2   & 0.4 \\

\hline
$\alpha$-elements
 &[$\alpha$/Fe]& 0.5 &  0.4   &  0.1   &  0.3   &  0.2   & 0.4    & 0.3 \\

\hline
s-process
 &[s/Fe]    & 1.5    & 2.2    & 1.5    & 1.1    & 0.9    & 1.6    & 0.0 \\
elements
 &[ls/Fe]   & 1.7    & 1.9    & 1.6    & 1.4    & 1.3    & 1.6    & 0.0 \\

 &[hs/Fe]   & 1.5    & 2.3    & 1.5    & 1.0    & 0.8    & 1.6    & 0.0 \\

\hline
\end{tabular}
\end{center}
\end{table*}

\subsection{Metallicity}

To calculate the metallicity, we used the iron abundance inferred from the FeII-lines since
this is by far the dominant ionization level.
The metallicity relative to the solar value ranges from $-$0.3 (\irtweetwee) to $-$1.0 (\irnulzeven).
This is a strong indication that our sample indeed consists of an old and 
hence low-mass population. This interpretation is further strengthened by 
the high galactic latitudes of the stars, in the range $|b|$=9$^{\circ}$--21$^{\circ}$.
An exception with respect to the latter criterion is \irtweedrie, for which $b=0.6^{\circ}$. Its 
low-mass nature is, however, clear from the low metallicity ([Fe/H]=$-$0.8).

\subsection{CNO abundance}

All stars display huge carbon-enrichments. This enrichment is beyond any
doubt even when the internal scatter of the C-lines and the uncertainty
induced by the model parameters are taken into account. As already pointed out in Sect.~\ref{subs:deter}, 
the O-abundance is in most cases less certain.
Only for the two hotter objects \irnulzeven\ and \ireennegen\ we feel
confident that we detected a moderate but real O-enrichment.
The reliability of the oxygen abundances will be
discussed in the following section, in which we treat the objects separately.

Unfortunately, the errors on mainly the O-abundance preclude accurate C/O number
ratio determination based on photospheric atomic lines. The range we determine goes
from 1.0 to 2.9.

\subsection{$\alpha$-elements}

The [$\alpha$/Fe] value is the mean of the [el/Fe] values of the observed
$\alpha$-elements. If we take into account a typical error of 0.2 dex in
the abundances, we may conclude that the $\alpha$-overabundances reflect
the chemical history of the Galaxy and are therefore not enhanced. 

The $\alpha$-elements show, however, large differences in their
[el/Fe] values within each star, up to an extreme value of 1 dex.
Unfortunately, the lack of useful lines for the light $\alpha$-elements
prevent too strong conclusions whether this is intrinisic or due to
non-LTE effects and/or unaccurate $log(gf)$.
The [$\alpha$/Fe] value is therefore only an indicative number. 

\subsection{s-process enrichment}

The s-process elements observed in evolved stars can be divided into two groups: the light s-process
elements around the magic neutron number 50 (Sr, Y, Zr) and the heavy
s-process elements around the magic neutron number 82 (Ba, La, Ce, Pr,
Nd, Sm, (Eu)). In order to investigate the s-process abundances in more
detail, four indexes are generally defined: [s/Fe], [ls/Fe], [hs/Fe] and [hs/ls].
Which elements are taken into account to determine these indices is different
from author to author and is mainly determined by the possibility to compute 
accurate abundances of the different elements. To compare the results on 
21\mic\, stars with other s-process enriched objects, the same index should be
used. We follow the proposition of Busso et al. (1995) and define the 
ls-index as the mean of Sr, Y and Zr and the hs-index as the mean of Ba, La,
Nd and Sm. Consequently, [s/Fe] is the mean of the 7 former elements and
[hs/ls]=[hs/Fe]$-$[ls/Fe].

For unobserved species of the light s-process no correction factor
was taken into account since the odd-even effect is not very strong.
For unobserved species of the hs index, for which the odd-even effect is much 
stronger, the undetermined abundance was estimated using 
the predictions of Malaney (1987) with a $\tau_{0}$ = 0.4 mbarn$^{-1}$ scaled
to the observed abundance of another heavy s-process element. This method
was used to estimate the Sm-abundance for \irnulvier\ and \ireennegen.
Note that the [hs/ls] index is by its definition in principle 
independent of the {\em total} s-process enrichment ([s/Fe]).

Regarding Table~\ref{tab:synopsis},
the overabundance of the s-process elements is beyond any doubt and is
definitely the most convincing argument for the post 3rd dredge-up status
of the studied stars. For unevolved objects in the same metallicity range,
one expects the s-process elements to scale with Fe: [s/Fe]=0 
(see Wheeler et al. 1989). 
Here we find [s/Fe]$\ge$0.9 (\irtweetwee), up to [s/Fe]=2.2 (\irnulvijf).

\begin{sidewaystable*}
\caption{Chemical analysis of the programme stars.  For every ion we list 
the solar abundance and for every star the number
of lines used (N), the mean equivalent width ($\overline{W_{\lambda}}$),
the abundance ratio relative to iron ([el/Fe]) and the internal scatter ($\sigma$
), if more than one line is used.
For the solar iron abundance we used the meteoric iron abundance of 7.51.
For the solar C, N and O abundances we adopted resp. 8.57, 7.99 and 8.86 
(C: Bi\'emont et al. 1993, N: Hibbert et al. 1991, O: Bi\'emont et al. 1991);
the other solar abundances are taken from Grevesse (1989). Note that abundances
based on one line should be treated with caution.}\label{tab:all}
\begin{center}
\begin{scriptsize}
\begin{tabular}{|ll|rrrr|rrrr|rrrr|rrrr|rrrr|rrrr|}
\hline
&&&&&&&&&&&&&&&&&&&&&&&&& \\

 &
 &
 \multicolumn{4}{|c|}{\bf \irnulvier } &
 \multicolumn{4}{|c|}{\bf \irnulvijf } &
 \multicolumn{4}{|c|}{\bf \irnulzeven} &
 \multicolumn{4}{|c|}{\bf \ireennegen} &
 \multicolumn{4}{|c|}{\bf \irtweetwee} &
 \multicolumn{4}{|c|}{\bf \irtweedrie} \\ 
 &
 &
 \multicolumn{4}{|c|}{\bf [Fe/H]=$-$0.62} &
 \multicolumn{4}{|c|}{\bf [Fe/H]=$-$0.85} &
 \multicolumn{4}{|c|}{\bf [Fe/H]=$-$1.00} &
 \multicolumn{4}{|c|}{\bf [Fe/H]=$-$0.60} &
 \multicolumn{4}{|c|}{\bf [Fe/H]=$-$0.31} &
 \multicolumn{4}{|c|}{\bf [Fe/H]=$-$0.79} \\ 
\hline 
&&&&&&&&&&&&&&&&&&&&&&&&& \\
 ion&
 Solar&
 N &$\overline{W_{\lambda}}$ & [el/Fe]& $\sigma$ &
 N &$\overline{W_{\lambda}}$ & [el/Fe]& $\sigma$ &
 N &$\overline{W_{\lambda}}$ & [el/Fe]& $\sigma$ &
 N &$\overline{W_{\lambda}}$ & [el/Fe]& $\sigma$ &
 N &$\overline{W_{\lambda}}$ & [el/Fe]& $\sigma$ &
 N &$\overline{W_{\lambda}}$ & [el/Fe]& $\sigma$ \\
&&&&&&&&&&&&&&&&&&&&&&&&& \\
\hline
&&&&&&&&&&&&&&&&&&&&&&&&& \\
 He\,I &10.99&   &    &       &     &   &   &       &      &      &   &       &  
   & 1 & 39 & +1.08 &      &   &   &       &      &    &   &       &     \\   
 C\,I  &8.57 & 7 & 87 & +0.76 & 0.20&11 & 96& +1.01 & 0.14 &   40 & 58& +1.08 & 0
.12&21 & 48& +1.01 & 0.12 & 8 & 70& +0.32 & 0.14 & 15 & 71& +0.92 & 0.12 \\ 
 N\,I  &7.99 & 2 & 65 & +0.39 & 0.01& 3 & 52& +0.69 & 0.08 &    8 & 84& +0.85 & 0
.15& 3 & 85& +0.98 & 0.11 & 4 & 54& +0.16 & 0.08 &  3 & 45& +0.48 & 0.13 \\  
 O\,I  &8.86 &   &    &       &     & 2 & 31& +0.56 & 0.00 &    8 & 44& +0.81 & 0
.11& 8 & 62& +0.70 & 0.08 & 3 & 24&$-$0.05& 0.04 &  4 & 15& +0.17 & 0.03 \\
 \hline                                                                          
Na\,I &6.33 &   &    &       &     & 1 & 26& +0.04 &      &   1  & 11& +0.46 &  
   & 1 & 10& +0.22 &      &   &   &       &      &  1 & 20& +0.08 &      \\
 Mg\,I &7.58 &   &    &       &     &   &    &       &     &   1  & 71& +0.06 &  
   & 3 & 95& +0.24 & 0.18 &   &   &       &      &    &   &       &     \\   
 Mg\,II&     &   &    &       &     &   &    &       &     &      &   &       &  
   & 4 & 48& +0.51 & 0.07 &   &   &       &      &    &   &       &     \\   
 Al\,I  &6.47 &   &    &       &     &   &    &       &     &     &   &       &  
   & 1 & 95& $-$0.28 &    & 4 & 16& $-$0.01&0.10 &    &   &       &     \\   
 Si\,I &7.55 & 3 & 55 & +0.79 & 0.26& 5 & 50& +0.59 & 0.16 &      &   &       &  
   &   &   &       &      & 14 & 46& +0.29 &0.14 & 5 & 39& +0.79 & 0.25 \\  
 S\,I  &7.21 & 1 & 57 & +0.43 &     & 1 & 42& +0.28 &      &    1 & 15& +0.40 &  
   & 1 &  6& +0.26 &      & 4 & 53& +0.04 & 0.19 & 4 & 36& +0.56 & 0.11 \\   
 Ca\,I &6.36 & 3 & 35 & +0.18 & 0.16& 6 & 55& +0.08 & 0.18 &    5 & 21&$-$0.11& 0
.08& 1 & 11& +0.49 &      & 6 & 57& $-$0.09&0.10 & 7 & 63& +0.29 & 0.18 \\
 Ca\,II&     &   &    &       &     &   &    &       &     &      &   &       &  
   &   &   &       &      & 1 & 16& $-$0.25 &    &     &   &       &     \\  
 Sc\,II&3.10 &   &    &       &     &   &    &       &     &    6 & 50& +0.22 & 0
.10& 6 & 49& +0.38 & 0.15 &   &   &         &    & 2 &110& +0.29 & 0.08 \\
 Ti\,I &4.99 &   &    &       &     &   &    &       &     &      &   &       &  
   &   &   &       &      & 2 & 26& +0.47 & 0.07 &    &   &       &     \\   
 Ti\,II&     &   &    &       &     & 1 & 52& +0.58 &      &   31 & 84& +0.11 & 0
.15&11 & 70& +0.19 & 0.15 &   &   &       &      & 3 &111& $-$0.18 & 0.15 \\ 
 V\,II &4.00 &   &    &       &     &   &    &       &     &    1 &107& $-$0.05& 
   & 1 & 97& +0.12 &      &   &   &       &      &    &   &       &     \\  
 Cr\,I &5.67 &   &    &       &     &   &    &       &     &    3 & 79& $-$0.15&0
.13& 3 & 50& +0.19 & 0.08 & 7 & 45& $-$0.08 &0.15&  1 &  5& +0.16 &     \\   
 Cr\,II&     &   &    &       &     &   &    &       &     &   18 & 58& +0.19 & 0
.13&20 & 51& +0.13 & 0.10 & 6 &118& $-$0.03 &0.23&  2 &40&$-$0.04 & 0.20 \\
 Mn\,I &5.39 &   &    &       &     &   &    &       &     &      &   &       &  
   &   &   &       &      & 1 & 38& $-$0.26 &    &    &   &       &     \\   
 Mn\,II&     &   &    &       &     &   &    &       &     &    2 & 22& +0.16 & 0
.05&   &   &       &      & 1 & 50& +0.13   &    & 1 &  6& $-$0.44 &     \\
 Fe\,I &7.51 & 8 & 32 & $-$0.04& 0.11&23 & 57& +0.05& 0.14 &   36 & 32&$-$0.08& 0
.15&15 &29& $-$0.05& 0.06 &41 & 57& $-$0.11 &0.14& 39 & 53& +0.01 & 0.22 \\  
 Fe\,II&     &10 & 74 &        & 0.10& 7 & 84&      & 0.14 &   17 & 67&     & 0.1
0  & 18 & 74&      & 0.06 &12 & 49&         &0.15&  7 & 63&       & 0.14 \\  
 Ni\,I &6.25 & 3 & 47 & +0.54  & 0.11& 7 & 45& +0.53& 0.19 &    2 &  7& $-$0.12&0
.11&    &   &      &      &19 & 35& $-$0.01 &0.16&  9 & 18& +0.14 & 0.09 \\ 
 Ni\,II&     &   &    &       &     &   &    &       &     &    1 & 31& $-$0.11 &
   &  4 & 27& +0.11& 0.15 &   &   &         &    &      &   &       &     \\ 
 Zn\,I &4.60 &   &    &       &     &   &    &       &     &    1 & 15& +0.37 &  
   &    &   &      &      &   &   &         &    &    &   &       &     \\   
 \hline                                                                          
Y\,II &2.24 & 3 & 84 & +1.98 & 0.06& 2 &129& +2.09 & 0.02 &   16 & 79& +1.50 & 0
.25&18 & 74& +1.34 & 0.18 & 4 &128& +1.51 & 0.09 & 2 & 51& +1.83 & 0.06 \\
 Zr\,II&2.60 & 4 & 33 & +1.34 & 0.23& 3 & 72& +1.75 & 0.15 &   19 & 59& +1.60 & 0
.22&22 & 80& +1.41 & 0.11 & 5 & 83& +1.04 & 0.15 & 4 & 33& +1.27 & 0.20 \\   
 Ba\,II&2.13 & 1 & 25 & +1.73 &     & 1 & 82& +2.43 &      &    1 & 53& +1.94 &  
   & 1 & 55& +1.48 &      & 1 & 13& +0.85 &      & 1 & 40& +2.03 &      \\   
 La\,II&1.22 & 5 & 59 & +1.74 & 0.13&17 & 94& +2.47 & 0.17 &   31 & 28& +1.65 & 0
.15& 6 & 26& +1.21 & 0.09 &16 & 68& +1.04 & 0.17 & 9 & 64& +1.88 & 0.20 \\   
 Ce\,II&1.55 & 2 & 24 & +1.20 & 0.04& 6 & 71& +2.40 & 0.14 &   50 & 43& +1.45 & 0
.21& 7 & 10& +0.91 & 0.13 & 8 & 52& +0.82 & 0.23 & 9 & 57& +1.54 & 0.19 \\   
 Pr\,II&0.71 &   &    &       &     &16 & 69& +2.40 & 0.19 &    4 & 10& +1.45 & 0
.11&   &   &       &      & 5 & 48& +0.92 & 0.24 & 4 & 29& +1.81 & 0.22 \\  
 Nd\,II&1.50 & 2 & 23 & +1.08 & 0.17&19 & 96& +2.47 & 0.19 &   29 & 19& +1.31 & 0
.12& 1 &  3& +0.52 &      &12 & 51& +0.73 & 0.20 &21 & 45& +1.67 & 0.20 \\   
 Sm\,II&1.00 &   &    &       &     & 4 & 71& +1.85 & 0.03 &    9 & 12& +1.04 & 0
.05&   &   &       &      &10 & 53& +0.50 & 0.20 & 5 & 21& +0.92 & 0.09 \\   
 Eu\,II&0.51 & 1 & 41 & +0.97 &     & 1 & 38& +1.22 &      &    1 & 30& +0.48 &  
   & 1 &  3& +0.26 &      & 2 & 47& +0.57 & 0.06 & 1 & 65& +1.21 &      \\  
 Hf\,II&0.88 &   &    &       &     &   &   &       &      &  1 & 83  & +1.84  &     & 1 & 12& +1.25 &      &   &   &       &      &     &   &       &     \\ 
\hline
 
\end{tabular}
\end{scriptsize}
\end{center}
\end{sidewaystable*}


\section{Individual objects}\label{sec:individobj}

The individual objects will be discussed in this section. 
In Table~\ref{tab:all} we list the complete
abundance analysis of each programme star. We want to stress that we will not
repeat the conclusions already made in the previous section, but that we
will focus on the peculiarities in the analysis of each star.

\subsection{IRAS\,04296+3429}

The chemical analysis of this object has already been discussed
by our group (Decin et al. 1998). Nevertheless, there are some minor differences
between the present analysis and the previous one, as a consequence
of an update of our line list. The main conclusions of Decin et al. (1998) remain valid.

\subsection{IRAS\,05341+0852}

This object was first mentioned by Geballe \& Van der Veen (1990) suggesting
that it is an evolved F-type supergiant displaying carbon-rich circumstellar
dust. Its 21\mic\ feature was discovered by Kwok et al. (1995). An extensive
abundance analysis was previously carried out by Reddy et al. (1997). They
found the star to be metal-poor, carbon-rich and s-process enriched.
We included this star in our sample to make the comparison with
other objects more significant by using the same line list.

For the model parameters, we found T$_\mathrm{eff}$=6500K, \\
$\log g$=1.0 and $\xi_t$=3.5 \kms.
The model parameters agree with
those found by Reddy (6500, 0.5, 5.0), but there are some 
differences in the abundances. In general, for species displaying a 
large difference, 
our analysis was based upon significantly more lines, which should 
imply that our results are intrinsically more reliable.

Besides the huge carbon enrichment
also oxygen is moderately enhanced ([O/Fe]=+0.6) but this value
is based only on two blended lines. 

Reddy et al. also found a significant overabundance of lithium and 
we confirm the presence
of the Li line at $\lambda$670.7nm in our spectra as well. 

The number of lines of the s-process elements is
impressive. A close look at the spectrum reveals that it is even completely
dominated by lines of s-process elements.  [s/Fe] (Table~\ref{tab:synopsis})
is 2.2, making \irnulvijf\ the most s-process enhanced 
post-AGB star known so far. Note the even-odd alternation of the
absolute abundances (Fig.~\ref{fig:malfiguur}) and the good internal consistency
($\sigma<0.20$). There is significant difference in the [el/Fe] ratios of
our analysis and the one of Reddy et al. (1997) reaching up to +0.8 dex.
Since we take much more and much weaker lines into account reaching a good
internal accuracy, and use the same lines as in the other programme stars,
we will use our abundances in the discussion.

\subsection{IRAS\,07134+1005 = HD\,56126}

With its 8.3 visual magnitude, this is the brightest star in our sample.
Considerable efforts have been performed already to reveal the properties of the 
circumstellar shell of this well-known post-AGB star. $^{12}$CO, 
$^{13}$CO, HCN and HCO$^+$ (Bujarrabal et al. 1992, Omont et al. 1993) 
have been detected in millimeter and radio line emission.
The 21\mic\ feature was discovered by Kwok et al. (1989) in its IRAS Low
Resolution Spectrum. Optical spectra revealed the presence of circumstellar C$_2$, 
$^{12}$CN and $^{13}$CN bands (Bakker et al. 1996; Bakker \& Lambert 1998a)
and $^{12}$C$^{13}$C and $^{12}$C$^{16}$O bands (Bakker \& Lambert 1998b)
, which led to an estimate for the $^{12}$C/$^{13}$C ratio : $^{12}$C/$^{13}$C=
72 $\pm$ 26. Mid-infrared (8-21\mic) imaging shows no spherical but 
possibly an axial symmetry of the dust emission (Meixner et al. 1997;
Dayal et al. 1998).

Concerning the {\em photosphere}, there is general agreement that \irnulzeven\ 
is a pulsating star, but different pulsation periods have been published
(e.g. L\`ebre et al. 1996). Limited abundance analyses have already been performed
by Parthasarathy et al. (1992) and by Klochkova (1995), showing chemical
evidence for the post-AGB status of this object. However, the high quality NTT-spectra
for this star permitted us to elaborate a much more detailed abundance analysis. 
We used 342 lines of 27 different ions.

A comparison with
the results of Klochkova shows good consistency of the model parameters,
the metallicity and the abundances of ions common in both analyses.
However, the latter author found a much larger abundance if deduced from
lines of neutral atoms than from lines of singly ionized atoms of the
same element. This was attributed to unknown non-LTE effects. We do not
confirm this result since careful analysis of our much higher S/N spectra
does not reveal the presence of most neutral lines, while we identified some 
lines differently. This illustrates that it is often very difficult to assign 
uncertain LTE abundance results of an individual object solely to non-LTE effects. Also the
effect of other uncertainties should be carefully studied and a more systematic approach using
data on several objects should be initiated to evaluate possible non-LTE
deviations of specific lines.

The ionization equilibrium of Ni is also fulfilled but for Cr we find a
$\Delta(\log A(\mbox{CrII})-\log A(\mbox{CrI}))=0.34$. A lower model gravity can
solve this inconsistency for Cr but it may also indicate non-LTE occupation numbers for
the different ions of Cr. We chose not to change our model parameters in order not 
to compromise significant comparison with the other stars in our sample.

A problem occurred when calculating the Fe abundance from lines of neutral iron with a very
low excitation potential: the lines with a very small lower excitation potential gave
systematically lower abundances. This is a strong indication that non-LTE over-excitation
occurs. We therefore discarded all FeI-lines with an excitation
potential $\chi<1.25$eV in our abundance determinations, also for all other programme stars.

The large line-to-line scatter in the abundance
of Y and Zr is solely caused by
only three lines of Y ($\lambda$578.169\,nm, \linebreak $\lambda$550.990\,nm and
$\lambda$528.982\,nm), and two lines of
Zr \linebreak ($\lambda$448.545\,nm and $\lambda$449.546\,nm).
These lines do not show any systematic trend
with another atomic parameter (wavelength, excitation potential, equivalent
width), so the a\-no\-ma\-lies are probably caused by inaccurate
$\log(gf)$-values. The only line in common with the other stars is the
Y\,II-line at 528.982\,nm in \irtweetwee. Also for this star, the
abundance inferred from this line is significantly lower than the 
abundances inferred from the other lines.

We also detected several lines of singly ionised
Hafnium (Z=72) in the blue part of the spectrum of \linebreak  \irnulzeven, but
only two of them turned out to be unblended
($\lambda$409.316\,nm,  $W_{\lambda}=83$\,m\AA\ 
and $\lambda$466.414\,nm,  $W_{\lambda}=19$\,m\AA).
Unfortunately, the Hf-abundance derived from these two lines 
differs by 0.4 dex  ($\log A({\rm Hf})=1.72$ and 2.09 respectively).
These abundances are both very high if compared with the calculations
of Malaney (1987). We expect $\log A({\rm Hf})\approx1.1$ for a 
neutron exposure of $\tau=0.3$\,mb$^{-1}$
(neutron exposure: see Sect.~\ref{section:neutronexp}).
No other possible identification could, however, be found.
An interpretation of this result is difficult,
not only because of the probably inaccurate $\log(gf)$-values, but also
because of the unknown chemical history of this r-process element.

\subsection{IRAS\,19500$-$1709 = HD\,187885}

This post-AGB candidate shows only a weak 21\mic\ feature (Justtanont et 
al. 1996) which is probably an effect of the higher temperature of this
object. We already
published an abundance analysis of this object (Van Winckel et al. 1996a)
based on ESO La Silla spectra. The latter spectra show He\,I lines at 
447.08\,nm and 587.6\,nm pointing to a hot source. Indeed, demanding excitation 
equilibrium of the Fe\,I-lines yielded a temperature T$_\mathrm{eff}$=8000K.

In this study we present a new analysis (Table~\ref{tab:all}) based 
on UES-spectra with a higher S/N and a broader spectral coverage. The analysis of the 
iron lines yielded the same atmospheric parameters, only the $\log(W_{\lambda}/\lambda)$-abundance
diagram points to a larger microturbulent velocity $\xi_t=6$\kms. The lines of the
the s-process elements Y and Zr favour a slightly lower value of $\xi_t\le5$\kms.
We again keep the atmospheric parameters obtained using the iron lines.
A change in the microturbulent velocity of $\sim$1\kms\ will only influence
the abundance deduced from the largest lines ($W_{\lambda}\ge120$m\AA) with
a typical value of 0.1 dex.

Comparing this analysis with our previous one, there are some quantitative 
differences, but our conclusions for this star remain unchanged. The s-process 
overabundance is now much better established, with more lines for the
ls-elements and new abundances for the hs-elements La, Ce, Nd and Eu.

We report also for this star the detection of a line of singly
ionised Hafnium at $\lambda$409.316\,nm. An equivalent width $W_{\lambda}=12\,$m\AA\  
gives an abundance of $\log A({\rm Hf})=1.53$ which is again quite
high if compared with the calculations of Malaney (1987). We expect
$\log A({\rm Hf})\approx1.0$ for a $\tau=0.2$\,mb$^{-1}$. 
Another argument against such a high
Hf-abundance is the non-detection of the Hf\,II line at $\lambda$369.973\,nm.
Using the obtained abundance and the $\log(gf)$-values of the VALD2 database,
this line should have an equivalent
width of $W_{\lambda}=9\,$m\AA. Inaccurate \hfill  $\log(gf)$-values can be
the cause of this inconsistency. Other possibly detectable Hf\,II lines
are all heavily blended.

\subsection{IRAS\,22223+4327}

The chemical composition of this star was recently 
discussed by Decin et al. (1998). We list the individual abundances 
here again since minor differences
are obtained due to an upgrate of the line-list and a homogenization of
the atomic data.

\subsection{IRAS\,23304+6147}

Kwok et al. (1989) classified this object as a proto-pla\-ne\-ta\-ry nebula
and discovered its 21\mic\ feature. The circumstellar dust turned out to be
C-rich as confirmed by several authors: detection of C$_2$, C$_3$ 
(Hrivnak 1995), CO, HCN (Omont et al. 1993), CN (Bakker et al. 1997) and
non-detection of OH (Likkel 1989). Hrivnak et al. (1999) found the object
to be resolved in V with an extent of about 2 arcsec.
Reddy \& Parthasarathy (1996) derived model 
parameters for this object by fitting a Kurucz model atmosphere through
the observed spectral energy distribution. This technique yielded 
T$_\mathrm{eff}$=5000K and $\log g$=1.0. Comparing our
spectra with the other stars in our sample, we concluded that this 
temperature was probably too low : the high excitation O-triplet around $\lambda$615.6nm is 
clearly detected (see Fig.~\ref{fig:oxy}). 
Indeed, also our spectroscopic determination
of the temperature using the iron-lines resulted in a T$_\mathrm{eff}$=6750K. The abundance analysis 
makes \irtweedrie\ join the group of post-AGB stars that 
clearly display chemical evidence of their post 3rd dredge-up character.

To our knowledge, this is the first detailed photospheric abundance study of this star.
The internal scatter of the four O-lines
is small indicating that the very high C/O ratio of 2.9 is probably real.
For the metallicity, we find the value [Fe/H]=$-$0.8.
The other iron peak elements do follow this deficiency.
Again, the s-process abundances are quite impressive for this star.
Note that the internal scatter of the different abundance determinations are somewhat
higher for this star than for the other programme stars as a consequence of the lower S/N of our
spectra (m$_{v}$=13.1).

\section{Neutron exposure of 21\mic\, stars}\label{section:neutronexp}

\subsection{Goodness-of-fit}

A theoretical parameter describing the neutron irradiation is the neutron
exposure rate $\tau$, which is defined as $\int_0^t N_n(t')V(t')dt'$ where
$N_n$ is the neutron density, and $V$ is the relative velocity of the neutrons relative to 
the seed nuclei. The thermal pulses on the AGB are normally modelled by
an exponential irradiation parameterized by a mean neutron exposure $\tau_0$ 
(e.g. Ulrich 1973). In most calculations, the neutron density and the
fraction of the inter-shell material that remains exposed to neutrons are adopted as constants. 
The efficiency of s-processing is then indicated by one parameter $\tau_0$.

\begin{figure}
\resizebox{\hsize}{!}{\includegraphics{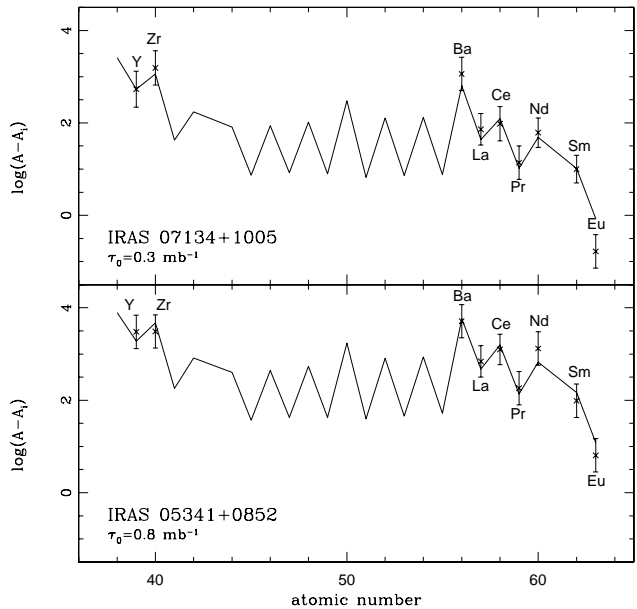}}
\caption{\label{fig:malfiguur}
Comparison of the observed abundances with the calculations
of Malaney (1987) for \irnulzeven\ and \irnulvijf. The observed abundances 
were corrected for the initial abundances (A$_i$) assuming [s/Fe]$_i$=0 
(except for Eu). 
As evidenced by the 'goodness of fit' procedure,
the s-process elemental distribution of \irnulzeven\ points
to a rather moderate neutron exposure, whereas the distribution of
\irnulvijf\ is best fitted by a large neutron exposure 
$\tau_0=0.8\,{\rm mb}^{-1}$.
}
\end{figure}

The neutron exposure $\tau_0$ is intimately related to the [hs/ls] index: 
a large $\tau_0$ yields a large [hs/ls]. In order to estimate
$\tau_0$ quantitatively, we used the theoretical models of Malaney (1987)
and the 'goodness of fit' procedure as defined by Cowley \& Downs (1980).
A description of this method can also be found in Vanture (1992) and
Smith et al. (1996). In this method, the quantity $S^2$ is defined as
 
$$
S^2=\frac{1}{N}\sum_{i=1}^N\frac{(O_i-(M_i+\delta))^2}{\sigma_i^2}
$$
where $N$ is the number of s-process elements involved in the comparison, $O_i$ the
observed abundance of element $i$, $M_i$ the predicted model abundance as
tabulated by Malaney, $\delta$ the average offset between the observed
and the predicted s-process abundances ($\delta=1/N\sum_i^N(O_i-M_i)$) and $\sigma_i$
is the uncertainty on the obtained individual abundance $O_i$. $O_i$ and $\sigma_i$ are
calculated as described in Decin et al. (1998) so on top of the internal accuracy a
fixed value of 0.3 was quadratically added to account for the uncertainties on the model
atmosphere parameters and eventual systematic offsets in the log(gf) values.
The values for $S^2$ for each model and each star are
listed in Table~\ref{tab:gof}. A graphical presentation of the
best fit for two stars (\irnulzeven\ and \irnulvijf) can be found
in Fig.~\ref{fig:malfiguur}. Note that the S$^{2}$ are small which indicates a
good consistency between the abundances of different s-process elements (see Smith, 1984).

 \begin{table}
 \begin{center}
 \caption{$S^2$ values for different neutron exposures $\tau_0$ (N$_n=10^{8}$cm$^{-3}$). The stars are 
listed with increasing [hs/ls]-value.\label{tab:gof}}
 \begin{tabular}{|l|rrrrr||r|}
 \hline
             & \multicolumn{5}{|c||}{$\tau_0$ (mb$^{-1}$)}  &[hs/ls] \\
             &0.1&0.2&0.3&0.4&0.5&        \\ 
             &0.6&0.7&0.8&0.9&1.0&        \\ 
 
 \hline
\irtweetwee & 1.55 & {\bf 0.45} & 0.55 & 0.76 & 0.95 &       \\
            & 1.11 & 1.23       & 1.35 & 1.44 & 1.51 & $-$0.5\\
\hline
\ireennegen & 1.29 &{\bf 0.34}  & 0.70 & 1.09 & 1.42 &       \\
            & 1.65 & 1.85 & 2.03       & 2.17 & 2.27 & $-$0.4\\
\hline
\irnulvier  & 2.35 & 0.80       &{\bf 0.74}&0.89&1.05&       \\
            & 1.18 & 1.28       &1.39  &1.48  &1.54  &$-$0.2 \\
\hline
\irnulzeven & 1.95 & 0.61       &{\bf 0.58}&0.73&0.88&       \\
            & 0.99 & 1.10 & 1.19       &1.26  &1.32  &$-$0.1 \\
\hline
\irtweedrie &3.36  &1.17        &0.76 &{\bf 0.72}&0.75&      \\
            &0.78  &0.82        &0.87 &0.91 &0.94    & +0.1  \\
\hline
\irnulvijf  &4.09  &1.28        &0.60 &0.39 &0.31    &       \\
            &0.29  &{\bf 0.28}  &{\bf 0.28}&0.29&0.30& +0.4  \\
 \hline
 \end{tabular}
 \end{center}
 \end{table}

The connection between $\tau_0$ and [hs/ls] is clear. Note also that for
the stars with a low [hs/ls] the minimum of $S^2$ and so the neutron 
exposure $\tau_0$ is well defined. For stars with a higher [hs/ls], a larger
range of values for $\tau_0$ are possible: a high neutron exposure
will only raise the overall level of enhancement without significantly
changing the distribution of the individual abundances (see also Fig. \ref{overabundance}). 
Therefore, the s-process distribution of \irnulvijf\ fits well for every 
$\tau_0 \ge 0.5\,{\rm mb}^{-1}$.


\subsection{Total s-process enrichment of 21\mic\, stars. }

\begin{figure}
\resizebox{\hsize}{!}{\includegraphics{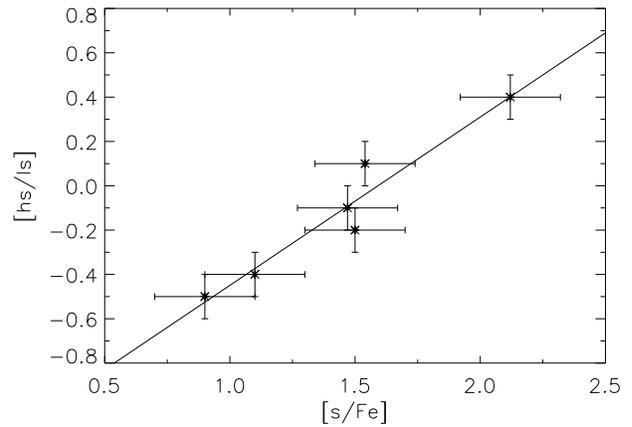}}
\caption{\label{fig:sfehsls}
The correlation between the total enrichment of s-process elements ([s/Fe] which
is the mean of [Y/Fe], [Zr/Fe], [Ba/Fe], [La/Fe], [Nd/Fe] and [Sm/Fe])
and the [hs/ls] index. The straight line gives the simple least-squares fit.
 }
\end{figure}

In Fig.~\ref{fig:sfehsls} the [hs/ls] index from the 21\mic\, stars is shown
in comparison with the overall s-process enrichment as parameterised 
by the [s/Fe]-index. More objects should clearly be studied, but the total neutron exposure 
is surprisingly strongly correlated with the total enrichment of s-process elements 
in the 21\mic\, stars, with more enriched objects showing also a stronger total neutron 
irradiation! A simple linear least-squares fit gives a high correlation coefficient of +0.96 with the fit 

$$
[\mathrm{hs}/\mathrm{ls}] = 0.76 [\mathrm{s}/\mathrm{Fe}] - 1.21 
$$

Since the asymptotic distribution 
of s-process elements is probably reached in carbon stars (Busso et al. 1995), 
the [hs/ls] index will not change with increasing \linebreak dredge-up of enriched material in the stellar 
envelope, making the [hs/ls] index independent of the dilution factor of the enriched 
material by the initial stellar envelope material. The [hs/ls] correlation with [s/Fe] therefore
reflects a different and more efficient internal nucleosynthesis, with
increasing dredge-up efficiency ! In AGB evolutionary models, the inclusion of protons into the intershell 
will create a primary $^{13}$C pocket and induce the neutron production.
Fig.~\ref{fig:sfehsls} indicates therefore that, with increasing dredge-up efficiency, also
the dredge-in of protons will increase, giving rise to a higher neutron irradiation.

\subsection{ Metallicity.}

Although our sample at this point only consists of 6 objects, a
considerable spread in metallicity ranging from
$-$0.3 to $-$1.0 is observed.
Since the $^{13}$C($\alpha$,n)$^{16}$O neutron source in AGB stars is most likely of primary
origin the [hs/ls] index should increase with decreasing metallicity due to the larger amount of 
neutrons per iron seed nucleus (e.g. Clayton 1988). This assumes, however, that the diffusion of protons 
into the C-rich layer itself is independent of metallicity or 
on parameters linked to the metallicity (like temperature profile, luminosity etc.).
A higher neutron exposure is therefore expected for low metallicity objects {\sl assuming} that 
the thermal pulse phenomenon itself is independent of metallicity.

\begin{figure}
\resizebox{\hsize}{!}{\includegraphics{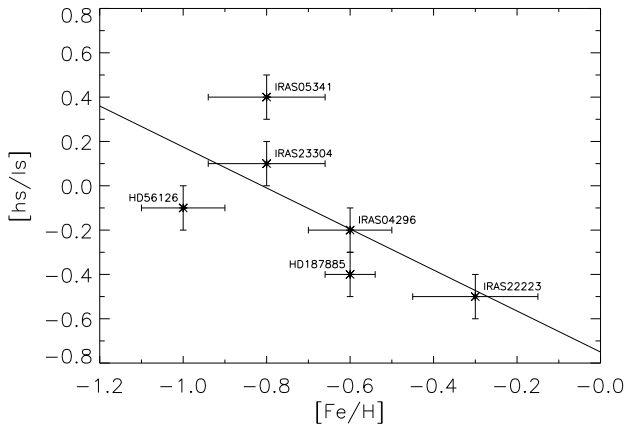}}
\caption{\label{fehslsa}
The [hs/ls] index of the analysed 21\mic\, stars as a function of the metallicity determined
by the Fe abundance ([Fe/H].  The line gives the simple least-squares fit 
[hs/ls] = -0.93 [Fe/H] $-$ 0.75.}
\end{figure}

In Fig.~\ref{fehslsa} we display the [hs/ls]-index in function of the metallicity. 
The 21\mic\, stars clearly cover a wide range in neutron
irradiation in the range of the observed iron abundances with indeed a inverse
dependence of the [hs/ls] ratio on metallicity. The correlation is, however, 
broad with a large intrinsic spread (the simple correlation coefficient is $-$0.67). The most 
metal deficient object \irnulzeven\ has seem to have undergone a much smaller neutron irradiation 
than the less metal deficient \irnulvijf. Clearly, also other stellar parameters 
determine the internal AGB nucleosynthesis during the AGB evolution.

Note that the two stars significantly below the least-square line are the ones for which
we detected the very heavy element Hf (Z=72). If indeed these high abundances turn out to be
true, this would indicate that the s-process is able to produce elements heavier than the Ba peak and 
thus probably up to the lead-peak! If true, the low [hs/ls]-index of the two stars might be
due to the production of very heavy elements. It is clear that high S/N blue spectra of 
all the objects are needed for a more detailed discussion.

Moreover, several (probably single) post-AGB objects are known which do not show s-process 
enrichment nor enhanced C/O ratios. These objects include HD\,161796 
([Fe/H]=$-$0.3; 
Luck et al., 1990), HD\,133656 ([Fe/H]= $-$1.0; Van Winckel et al. 1996b) 
and SAO\,239853 ([Fe/H]= $-$0.8, Van Winckel 1997) which show
very similar metallicities and IR-excesses. Unfortunately, like the 21\mic\, objects they are too 
far for reliable HIPPARCOS parallaxes so the luminosity differences between the enriched and 
non-enriched objects is an as yet unknown but probable key parameter in this discussion.


\section{Comparison with intrinsic and extrinsic enriched stars.}

\subsection{Intrinsic enriched objects: AGB stars}

In this section we compare our results on 21\mic\, sources
with the observational data on the intrinsic AGB stars of the M-MS-S-MC-C sequence. 
For MS and S stars we used results from the series of papers by Smith \& Lambert (1985,1986,1990)
and Smith et al. (1987) : since the spectra of oxygen rich AGB stars are dominated by strong 
TiO-bands, only a few spectral windows can be used to infer abundances from atomic lines. 
The regions in between
740 and 758\,nm, and 998 to 1010\,nm, were used together with additional IR spectra
around 1.6, 2.2 and 4.0 \mic\, to deduce abundances of CNO isotopes.
The situation is even worse for C stars where the C$_{2}$ and CN bands dominate the optical and
IR spectrum. To our knowledge only Utsumi (1985 and references therein) made a quantitative
analysis of the s-process enhancement in C stars using two small spectral windows with minimal
molecular absorption (475--490\,nm, and 440--450\,nm).
SC stars, on the other hand, are AGB stars with a C/O ratio close to 1.0 making the oxide bands and bands from
carbon molecules very weak or absent so that many atomic lines become detectable in the
spectrum. We use the results for seven SC stars from a recent paper by 
Abia \& Wallerstein (1998).

\begin{figure}
\resizebox{\hsize}{!}{\includegraphics{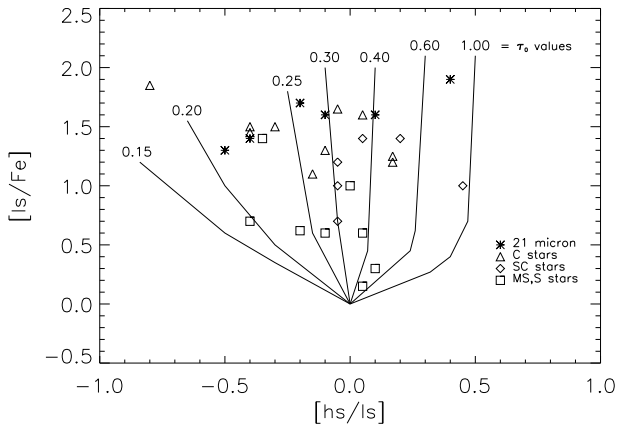}}
\caption{\label{overabundance}
Comparison between the overabundances of light s-process elements 
observed in the 21\mic\, stars 
with those from intrinsic AGB stars. 'ls' is the mean of the light s-process 
elements Y and Zr, while
'hs' is the mean of the heavy s-process elements Ba, La, Nd and Sm. 
The references for the observations of the intrinsic AGB stars can be found 
in Abia \& Wallerstein (1998).
The full lines are the theoretical predictions for different neutron exposures 
parameterised by $\tau_{0}$\, (in mbarn$^{-1}$) are from Busso et al. (1995).
 }
\end{figure}

In Fig.~\ref{overabundance} we compare the observed overabundances of the light s-process elements
([ls/Fe]) with the ones observed in intrinsic AGB stars. Since there is strong observational evidence
that the 21\mic\, stars are post-carbon stars, it is no surprise that the overabundances of
the s-process elements in 21\mic\, stars are indeed large in comparison with MS, intrinsic S and
SC stars and are comparable with and on average even slightly higher than observed in C stars. 
We therefore can confirm earlier results that the increasing C/O ratio in the MS-S-SC-C
sequence  is also reflected in an increasing enhancement of s-process elements 
(e.g. Smith \& Lambert 1990; Abia \& Wallerstein 1998).

\subsection{Extrinsic enriched objects}\label{sec:neut&met}

To investigate the metallicity dependence of the integrated neutron exposure for intrinsic AGB stars 
is not easy since the metallicity of field carbon stars is very difficult to determine
and the spread in metallicities observed
in Galactic M-MS-S-SC stars is not large (e.g. Smith \& Lambert 1990). 
Plez et al. (1993) have shown, however, that 
seven intrinsic AGB stars in the metal poor ([Fe/H] = $-$0.5) Small Magellanic Cloud do indeed show a
very high neutron exposure and they suggest that it is not intrinsic to the SMC but due to the
low initial metallicity.

The metallicity spread observed in the extrinsic enriched objects is much larger 
and there is growing observational evidence that the [hs/ls] index is indeed anti-correlated 
with metallicity (e.g. Smith 1999): 
Busso et al. (1995) confirmed earlier suggestions by Luck \& Bond (1991) in a compilation 
of literature 
values of the [hs/ls] index in Ba stars and MS-S stars and quantified tentatively the metallicity 
dependence of the neutron irradiation as an increase in [hs/ls] of about 0.2 dex for a metallicity 
drop from 0 to $-$0.5. Strong support came
also from the work by Vanture (1992) who has shown that the neutron irradiation is high in
the metal-poor CH giants, thus enlarging the metallicity spread to lower metallicities. 
North et al. (1994)
also found that the [hs/ls] index was correlated with the metallicity in a sample of CH-subgiants.
Finally also the low metallicity yellow symbiotics, who recently were included in the Ba-star family,
show s-process distributions characterised by a relatively high neutron irradiation 
(Smith et al. 1996, 1997;
Pereira et al. 1998). The results obtained by Luck \& Bond (1991) included, however, 4 metal 
deficient objects with small overabundances showing a very weak neutron irradiation. They 
dubbed these stars ``the metal-deficient Ba stars''. 

Some caution should be expressed by generalising these results to single star AGB evolution :
especially for the short orbital period binaries there is strong evidence that the binarity
affects the mass-loss and thus the AGB evolution itself of these binaries. The amount of thermal
pulses and the characteristics of the s-process synthesis might therefore also be affected by
the orbital elements (e.g. Jorissen \& Mayor 1988).

\begin{figure}
\resizebox{\hsize}{!}{\includegraphics{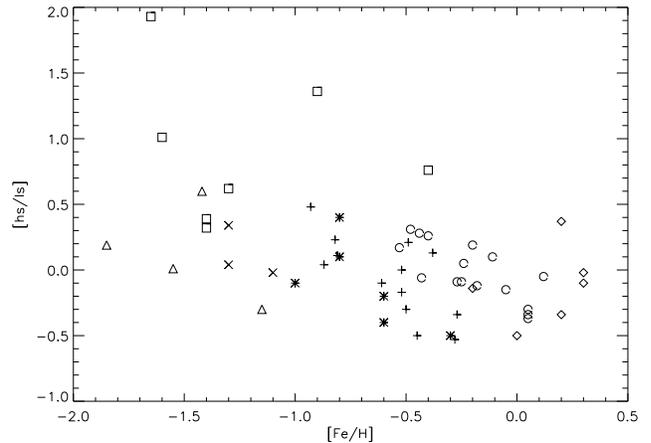}}
\caption{\label{fehslsb}
The [hs/ls] index as a function of the metallicity for a sample
of s-process enriched objects.  Asterisks are the 21\mic\, stars; triangles the 
metal deficient Barium stars from Luck \& Bond 1991; squares the CH-giants 
from Vanture (1992);
circles the Ba-stars from the compilation made by Busso et al. (1995); plus-signs the
CH-subgiants from North et al. (1994); crosses the yellow symbiotic stars
from Smith et al. (1996,1997) and Pereira et al. (1998); diamonds the SC stars form
Abia \& Wallerstein (1998).
}
\end{figure}

In Fig.~\ref{fehslsb} we compare the [hs/ls] index in function of the metallicity 
with the values found in the literature for other s-process enriched objects. 
This compilation is difficult since the definition of the hs and ls indices is
author-dependent. We scaled the abundances to the definition given in section 4.4. For
undetermined abundances of the CH giants, we used the tables of Malaney (1987) with a
$\tau_{0}$ of 1.0 mbarn$^{-1}$ which is more appropriate for those objects, while for the others we 
used 0.4 mbarn$^{-1}$ tables. We discarded
objects for which less than 4 abundances of the in total 7 elements (Sr, Y, Zr, Ba, La, Nd and
Sm) were available.

Fig.~\ref{fehslsb} shows that there is a large spread in the integrated neutron irradiation, but that
there is indeed a global increase with decreasing metallicities from [Fe/H] $\sim$ $-$0.5 downwards as
already noted by several authors (e.g. Smith, 1999). 
Three out of four ``metal-deficient Ba stars'' show a very low [hs/ls]-index (Luck $\&$ Bond, 1991). 
These stars certainly are worth a re-analysis based on higher resolution, higher 
signal-to-noise spectra since the internal errors of the analysis based on photographic intermediate
resolution (0.2 \AA) spectra are rather large and the observed overabundances small 
(Luck \& Bond, 1991). Some CH-stars show extremely high [hs/ls]-indices for their 
metallicity but also some Ba stars
like the very Nd-rich HD\,35155 (not in the figure) show a-typical values. The 21\mic\, post-AGB 
objects follow the same trend as the CH-subgiants of the same metallicity and show 
on average a somewhat lower [hs/ls]-index compared to the Ba giants.

We can conclude that the metallicity dependence of the neutron exposure has a large {\sl intrinsic}
spread which confirms that certainly also other fundamental parameters determine the internal 
nucleosynthesis and dredge-up phenomena during the AGB evolution.


\section{Conclusions}

We presented a homogeneous study of abundances displayed by a sample of
carbon-rich 21\mic\, post-AGB stars. With their low metallicity ([Fe/H] ranging from $-$0.3 to $-$1.0),
high C/O ratio (ranging from 1.0 to 2.9) and especially their high photospheric abundance
of s-process elements, we confirm that they are post-3rd-dredge-up objects. 
Since also the circumstellar environment is C-rich and no orbital motion is observed we can 
classify the 21\mic\, stars as single post-carbon stars of low initial mass.

The G--F spectral type makes that a wide variety of s-process elements
can be detected by their atomic lines. This, together with their significant spread in metallicity,
make that the 21\mic\, stars are ideal objects to constrain the AGB nucleosynthesis and 
dredge-up models. The most interesting constraint in this context comes from the possibility to 
accurately determine a wide variety of elemental abundances of these intrisically enriched objects. 
Contrary to AGB stars themselves, no photospheric 
molecular absorption is observed, so isotopic 
ratios of key elements are in the 21\mic\, stars unfortunately difficult to determine.
The molecular bands originating in the circumstellar environment
can, however, be used for some isotopes like $^{12}$C and $^{13}$C (e.g. Bakker \& Lambert 1998a).

We can summarise the main conclusions with respect to the s-process nucleosynthesis 
displayed by 21\mic\, stars as follows :

\begin{itemize}

\item The 21\mic\, stars display a wealth of atomic lines from s-process elements. The photospheric
spectrum of stars like \irnulvijf\ is even completely dominated by atomic lines of heavy elements.
The observed overabundances of s-process elements in the analysed 21\mic\, stars are large and 
range from [ls/Fe] = $+$1.3 to $+$1.9 which make the 21\mic\, stars among the most
s-process enriched objects known.

\item The 21\mic\, stars display a wide variety of integrated neutron 
irradiation as parameterised by the observational 
[hs/ls]-index, corresponding to a $\tau_{0}$ neutron exposure from $\tau_{0}$ = 0.2 to 0.8 mbarn$^{-1}$.

\item The [hs/ls] index displayed by the 21\mic\, stars is
strongly correlated (classical correlation coefficient of +0.96) with the total enrichment of 
the s-process elements as parameterised by the [s/Fe]-index, in the sense that more enriched objects
also display a higher integrated neutron irradiation. Since in carbon stars the asymptotic values of
the s-process distribution is probably reached, this means that the dredge-up efficiency is strongly 
linked with the neutron production in the intershell !

\item The anti correlation of the [hs/ls] index with the metallicity of the 21\mic\, objects 
is less well determined (classical correlation coefficient $-$0.67) and contains a large scatter.
This scatter is certainly intrinsic and confirms that also other fundamental parameters
determine the internal nucleosynthesis and dredge-up phenomena during the AGB evolution.
Note that the 21\mic\ stars are the only intrinsically enriched objects for which a wide
range in metallicity is observed.

\item Compared with other s-process enriched objects of extrinsic
nature, the neutron exposure displayed by the 21\mic\,stars fall roughly in the same range
as the CH subgiants and is on average lower than it is in Ba-giants of the same
metallicity. Note that the metallicity dependence of the
neutron exposure displayed by intrinsic and extrinsic objects contains also a large intrinsic scatter. 
Especially the iron deficient Ba stars as defined
by Luck \& Bond (1991) should be analysed in great detail since
they display a very low neutron irradiation for their low metallicity.

\end{itemize}

We can conclude that a homogeneous detailed study of the 21\mic\ stars forms a very 
useful complement to other types
of s-process enriched objects (intrinsic AGB stars and extrinsic enriched objects) used in the 
observational study of 
the theoretically less well understood s-process nucleosynthesis, mass-loss and dredge-up phenomena 
occurring on the AGB.

\begin{acknowledgements}
The authors would like to thank the staff of the NTT and WHT telescopes for
the welcome support during the observations. 
It is a pleasure to acknowledge the Vienna Atomic Line Database (VALD2) for the
usefull data-set and Prof. Kurucz for the distribution of his software. Also
Eric Bakker is warmly acknowledged for providing some of the spectra as well
as Christoffel Waelkens and Nami Mowlavi for stimulating discussions.
HVW and MR acknowledge support from the Fund of Scientific Research, Flanders.

\end{acknowledgements}


\begin{thebibliography}{}

\bibitem[]{}
Abia C., Wallerstein G., 1998, \mnras{293}, 89 
\bibitem[]{}
Bakker E.J., Lambert D.L., 1998a, \apj{502}, 417 
\bibitem[]{}
Bakker E.J., Lambert D.L., 1998b, \apj{508}, 387 
\bibitem[]{} 
Bakker E.J., Waters L.B.F.M., Lamers H.J.G.L.M., et al., 1996, \aea{310}, 893
\bibitem[]{} 
Bakker E.J., Van Dishoeck E.F., Waters L.B.F.M., Schoenmaker T., 1997, 
\aea{323}, 469
\bibitem[]{}
Biehl D., 1976, PhD thesis, Univ. Kiel
\bibitem[]{} 
Bi\'emont E., Hibbert A., Godefroid M., Vaeck N., Fawett B.C., 1991, \apj{375}, 818
\bibitem[]{}
Bi\'emont E., Hibbert A., Godefroid M., Vaeck N., 1993, \apj{412}, 431
\bibitem[]{}
Blackwell D.E., Shallis M.J., Simmons G.J., 1980, \aea{81}, 340 
\bibitem[]{}
Bujarrabal V., Alcolea J., Planesas P., 1992, \aea{257}, 701 
\bibitem[]{}
Busso M., Gallino R., Lambert D.L., Raiteri C.M., Smith V.V., 1992, \apj{399},
218
\bibitem[]{}
Busso M., Lambert D.L., Beglio L., et al., 1995, \apj{446}, 775   
\bibitem[]{}
Clayton D.D., 1988, \mnras{234}, 1
\bibitem[]{}
Cowley C.R., Downs P.L., 1980, \apj{236}, 648
\bibitem[]{} 
Dayal A., Hoffmann W.F., Bieging J.H., et al., 1998, \apj{492}, 603 
\bibitem[]{}
Decin L., Van Winckel H., Waelkens C., Bakker E.J., 1998, \aea{332}, 928 
\bibitem[]{}
Edvardsson B., Andersen J., Gustafsson B., et al., 1993, \aea{275}, 101
\bibitem[]{}
Frost C.A., Lattanzio J.C., 1996, \apj{473}, 383
\bibitem[]{}
Fuhr J.R., Martin G.A., Wiese W.L., 1988, Journal
of Physical and Chemical Reference Data, Volume 17, Supplement No. 4
\bibitem[]{}
Gallino R., Arlandini C., Busso M., et al., 1998, \apj{497}, 38
\bibitem[]{}
Geballe T.R., Van Der Veen W.E.C.J., 1990, \aea{235}, L9 
\bibitem[]{} 
Gonzalez G., Lambert L., Giridhar S., 1997, \apj{481}, 452
\bibitem[]{} 
Grevesse N., 1989, in AIP Conferences Series 183 : ``Cosmoc Abundances of
Matter'', ed C.J. Waddington, American Institute of Physics, New York, p.9 
\bibitem[]{}
Herwig F., Bl\"{o}cker T., Sch\"onberner D., El Eid M., 1997, \aea{324}, L81
\bibitem[]{} 
Hibbert A., Bi\'emont E., Godefroid M., Vaeck N., 1991, A\&ASS 88, 505
\bibitem[]{} 
Hibbert A., Bi\'emont E., Godefroid M., Vaeck N., 1993, A\&ASS 99, 179
\bibitem[]{}
Hrivnak B.J., 1995, \apj{438}, 341  
\bibitem[]{}
Hrivnak B.J., Kwok S., 1999, \apj{513}, 869
\bibitem[]{}
Hrivnak B.J., Langill P.P., Su H.Y.L., Kwok S., 1999, ApJ 513, 421
\bibitem[]{}
Jorissen A., 1999, in : IAU Symp. 191
``AGB stars'', Le Bertre T., L\`{e}bre A., Waelkens C. (eds.), 437
\bibitem[]{}
Jorissen A., Arnould M., 1989, \aea{221}, 161
\bibitem[]{} 
Jorissen A., Mayor M., 1988, \aea{198}, 187
\bibitem[]{}
Jorissen A., Frayer D.T., Johnson H.R., Mayor M., Smith V.V., 1993, \aea{271}, 463
\bibitem[]{} 
Justtanont K., Barlow M.J., Skinner C.J., et al., 1996, \aea{309}, 612
\bibitem[]{}
Klochkova V.G., 1995, \mnras{272}, 710 
\bibitem[]{}
Klochkova V.G., Szczerba R., Panchuk V.E., Volk K., 
1999, \aea{345}, 905 
\bibitem[]{} 
Knapp G.R., Sutin B.M., Phillips T.G., et al., 1989, \apj{336}, 822 
\bibitem[]{}
Kupka F., Piskunov N.E., Ryabchikova T.A., Stempels H.C., Weiss W.W.,
1999, \aea{submitted}
\bibitem[]{} 
Kurucz R.L., 1993, CD-ROM set, Cambridge, MA : Smithsonian
Astrophysical Observatory
\bibitem[]{}
Kwok S., Volk, K.M., Hrivnak, B.J., 1989, \apj{345}, 51   
\bibitem[]{}
Kwok S., Hrivnak B.J., Geballe T.R., 1995, \apj{454}, 394 
\bibitem[]{}
Kwok S., Volk K., Hrivnak B.J., 1999, in : IAU Symp. 191 ``AGB stars'', Le Bertre T., L\`{e}bre A., 
Waelkens C. (eds.), 297
\bibitem[]{}
Lambert D.L., 1989, in AIP Conferences Series 183 : ``Cosmoc Abundances of
Matter'', ed C.J. Waddington, American Institute of Physics, New York, p.168
\bibitem[]{}
Lambert D.L., Smith V.V., Busso M., Gallino R., Straniero O., 1995, 
\apj{450}, 302
\bibitem[]{} 
Lambert D.L., Heath J.E., Lemke M., Drake J., 1996, \apj{103}, 183
\bibitem[]{}
Langer N., Heger A., Wellstein S., Herwig F., 1999, \aea{346}, L37
\bibitem[]{}
L\`ebre A., Mauron N., Gillet D., Barthes D., 1996, \aea{310}, 923 
\bibitem[]{}
Likkel L.,  1989, \apj{344}, 350 
\bibitem[]{}
Likkel L., Forveille T., Omont A., Morris M., 1991, \aea{246}, 153 
\bibitem[]{}
Luck R.E., Bond H.E., 1991, \apjss{77}, 515
\bibitem[]{}
Luck R.E., Bond H.E., Lambert D.L., 1990, \apj{357}, 188
\bibitem[]{}
Malaney R.A., 1987, Ap\&SS 137, 251
\bibitem[]{}
Martin G.A., Fuhr J.R., Wiese W.L., 1988, Journal
of Physical and Chemical Reference Data, Volume 17, Supplement No. 3
\bibitem[]{}
McWilliam A., 1997, \araa{35}, 503
\bibitem[]{}
McWilliam A., Preston G.W., Sneden C., Searle L., 1995, \aj{109}, 2757
\bibitem[]{} 
Meixner M., Skinner C.J., Graham J.R., et al., 1997, \apj{482}, 897
\bibitem[]{}
Mowlavi N., 1999, A\&A 344, 617
\bibitem[]{}
Mowlavi N., Jorissen A., Arnould M., 1998, \aea{311}, 803
\bibitem[]{}
North P., Berthet S., Lanz T., 1994, \aea{281}, 775 
\bibitem[]{}
Omont A., Loup C., Forveille T., et al., 1993, \aea{267}, 515 
\bibitem[]{} 
Oudmaijer R.D., Bakker E.J., 1994, \mnras{271}, 615 
\bibitem[]{}
Parthasarathy M., Garcia Lario P., Pottasch S.R., 1992, \aea{264}, 159 
\bibitem[]{}
Pereira C.B., Smith V.V., Cunha K., 1998, \aj{116}, 1977
\bibitem[]{}
Plez B., Smith V.V., Lambert D.L., 1993, \apj{418}, 812 
\bibitem[]{}
Reader J., Corliss C.H., Wiese W.L., Martin G.A., 1980, NSRDS-NBS 68, 1
\bibitem[]{}
Reddy B.E., Parthasarathy M., 1996, \aj{112}, 2053 
\bibitem[]{}
Reddy B.E., Parthasarathy M., Gonzalez G., Bakker E.J., 1997, \aea{328}, 331
\bibitem[]{}
Smith V.V., 1984, \aea{132}, 326
\bibitem[]{}
Smith V.V., 1999, in: IAU symp. 191 `` AGB stars '', Le Bertre T., L\`ebre A.,
Waelkens C. (eds.), ASP conference series, 69
\bibitem[]{}
Smith V.V., Lambert D.L., 1985, \apj{294}, 326
\bibitem[]{}
Smith V.V., Lambert D.L., 1986, \apj{311}, 843
\bibitem[]{}
Smith V.V., Lambert D.L., 1990, \apjss{72}, 387
\bibitem[]{}
Smith V.V., Lambert D.L., McWilliam A., 1987, \apj{320}, 862
\bibitem[]{}
Smith V.V., Cunha K., Jorissen A., Boffin H.M.J., 1996, \aea{315}, 179  
\bibitem[]{}
Smith V.V., Cunha K., Jorissen A., Boffin H.M.J., 1997, \aea{324}, 97
\bibitem[]{}
Straniero O., Gallino R., Busso M., et al., 1995, \apj{440}, L85
\bibitem[]{} 
Th\'{e}venin F., 1989, \aeass{77}, 137
\bibitem[]{}  
Th\'{e}venin F., 1990, \aeass{82}, 179
\bibitem[]{}
Ulrich R.K., 1973, in: D.N. Schramm \& W.S. Arnett (eds) Explosive
Nucleosynthesis. Austin. University of Texas Press, p. 139
\bibitem[]{}
Utsumi K., 1985, in : M. Jashek \& P.L. Keenan (eds) Cool Stars with
Excesses of Heavy Elements. Reidel, Dordrecht, p. 243 
\bibitem[]{}
Van Eck S., Jorissen A., Udry S., Mayor M., Pernier B., 1998, \aea{329}, 971
\bibitem[]{}
Vanture A.D., 1992, \aj{104}, 1997
\bibitem[]{} 
Van Winckel H., 1997, \aea{319}, 561 
\bibitem[]{}
Van Winckel H., 1999, in : IAU Symp. 191
``AGB stars'', Le Bertre T., L\`{e}bre A., Waelkens C. (eds.), 465
\bibitem[]{}
Van Winckel H., Waelkens C., Waters L.B.F.M., 1996a, \aea{306}, L37 
\bibitem[]{}
Van Winckel H., Oudmaijer R.D., Trams N.R., 1996b, \aea{312}, 553
\bibitem[]{} 
Venn K.A., 1995, \apjss{99}, 659
\bibitem[]{}
Volk K., Kwok S., Hrivnak B.J., 1999, \apj{516}, L99 
\bibitem[]{} 
Wheeler J.C., Sneden C., Truran J.W.Jr., 1989, \araa{27}, 279 
\bibitem[]{}
Wiese W.L., Smith M.W., Glennon B.M., 1966, NSRDS-NBS, Vol I, 4
\bibitem[]{}
Woodsworth A.W., Kwok S., Chan S.J., 1990, \aea{228}, 503 
\bibitem[]{}
Za\v{c}s L., Klochkova V.G., Panchuk V.E., 1995, \mnras{275}, 764 

\end{thebibliography}
\end{document}